\documentclass[]{aastex631}
\usepackage{bm}
\usepackage{amsmath}  
\usepackage{amssymb}  
\usepackage{multirow}
\usepackage{upgreek}

\newcommand{\msun}{{\rm M}_\odot}
\newcommand{\msunyr}{{\rm M}_\odot\,{\rm yr}^{-1}}

\newcommand{\cc}{{\rm cm^{-3}}}

\newcommand{\Jang}{{\rm g\,cm^{-2}\,s^{-1}}}
\newcommand{\kms}{{\rm km\,s^{-1}}}
\newcommand{\fj}{{\rm g\,cm^{-2}\,s^{-2}}}

\newcommand{\vect}[1]{\mbox{\boldmath$#1$}}

\shorttitle{Cloud Dissipation and Disk Wind} 
\shortauthors{Machida and Basu}

\begin{document}

\title{Cloud Dissipation and Disk Wind in the Late Phase of Star Formation}

\correspondingauthor{Masahiro N. Machida}
\email{machida.masahiro.018@m.kyushu-u.ac.jp}

\author[0000-0002-0963-0872]{Masahiro N. Machida}
\affiliation{Department of Earth and Planetary Sciences, Faculty of Science, Kyushu University, Fukuoka 819-0395, Japan}
\affiliation{Department of Physics and Astronomy, University of Western Ontario, London, ON N6A 3K7, Canada}

\author[0000-0003-0855-350X]{Shantanu Basu}
\affiliation{Department of Physics and Astronomy, University of Western Ontario, London, ON N6A 3K7, Canada}

\begin{abstract}
We perform a long-term simulation of star and disk formation using three-dimensional nonideal magnetohydrodynamics. The simulation starts from a prestellar cloud and proceeds through the
long-term evolution of the circumstellar disk until $\sim 1.5\times10^5$ yr after protostar formation.
The disk has size $\lesssim 50$ au and little substructure in the main accretion phase because of the action of magnetic braking and the magnetically-driven outflow to remove angular momentum. 
The main accretion phase ends when the outflow breaks out of the cloud, causing the envelope mass to decrease rapidly. 
The outflow subsequently weakens as the mass accretion rate also weakens.  
While the envelope-to-disk accretion continues, the disk grows gradually and develops transient spiral structures due to gravitational instability. 
When the envelope-to-disk accretion ends, 
the disk becomes stable and reaches a size $\gtrsim 300$\,au. In addition, 
about 30\% of the initial cloud mass has been ejected by the outflow. 
A significant finding of this work is that after the envelope dissipates, a revitalization of the wind occurs, and there is mass ejection from the disk surface that lasts until the end of the simulation.
This mass ejection (or disk wind) is generated since
the magnetic pressure significantly dominates both the ram pressure and thermal pressure above and below the disk at this stage.
Using the angular momentum flux and mass loss rate estimated from the disk wind, 
the disk dissipation timescale is estimated to be $\sim10^6$\,yr. 
\end{abstract}

\keywords   
{
Magnetohydrodynamical simulations (1966) ---
Protostars (1302) ---
Protoplanetary disks (1300) ---
Circumstellar disks (235) ---
Stellar jets (1807) ---
Star formation (1569) 
}

\section{Introduction}
\label{sec:intro}

The final phase of star and planet formation takes place
through the protoplanetary disk. Understanding the formation, evolution, and dispersal of these disks is key to understanding how planets are formed. In recent years, many new insights have been gained through high sensitivity and resolution observations of disks using the Atacama Large Millimeter/submillmeter Array (ALMA).  
Ring and gap structures have been clearly observed in protoplanetary disks around Class II objects \citep{andrews18,cieza19,oberg21}, and are considered to be important signposts of ongoing planet formation. In addition to being found around Class II objects, the Keplerian (or rotationally supported) disks have also been confirmed around Class 0 and I objects in ALMA observations \citep{hara13,murillo13,ohashi14,aso15,aso19,maret20,ohashi23}. Ring and gap structures were found in some disks around very young protostars, or Class 0/I objects \citep{sheehan17,sheehan18,sheehan20,shoshi24}, but very recent high-resolution observations find a lack of substructures in a sample of Class 0 disks \citep{ohashi23}. 

The evolution of isolated disks has often been attributed to angular momentum transport due to internal turbulence generated by the magnetorotational instability (MRI) \citep{Balbus98,balbus03,mcKee07}. It should be noted that since an isolated disk was considered in the classical picture of planet formation, the angular momentum transfer due to the outflow and magnetic braking on a large scale had been ignored in many studies. 
However, recent observations have indicated that the scale height of the protoplanetary disk is very small, implying  that turbulence caused by MRI is not sufficient to inflate the disk in the vertical direction \citep[e.g.,][]{pinte16,dullemond18,flaherty20,pizzati23}. Spectral line observations also show that the line broadening implies weak turbulence \citep{flaherty15}.
In such a case, it would be difficult to transport the angular momentum of the disk within the required timescale.
Observational estimates are that protoplanetary disks are largely dispersed within 10 Myr, with a typical lifetime of $2-3$ Myr \citep[e.g.,][]{williams11}.
In addition, recent theoretical studies show that it is difficult to induce MRI in the protoplanetary disk, because the ionization degree in the disk is very low and the dissipation of the magnetic field can suppress the growth of the MRI \citep{turner08,bai13,simon13a,simon13b,gressel15,bai15,kawasaki21}. As an alternative, disk winds have been suggested as the source of angular momentum removal that drives transport of material within the disk \citep{bai13,suzuki16}.
The disk winds around Class II objects have been confirmed in many observations \citep{zapata15,louvet18,devalon20,Fern20,booth21,launhardt23,fang23}.
The mass and density of the wind are expected to be very small after the main accretion phase, because the mass accretion onto the disk, which gives a power to drive the wind, is almost complete by this epoch. 
However, a non-negligible mass and angular momentum might be ejected from the young stellar system with the disk wind. 
Some analytical studies indicate that the angular momentum can be transferred by a wind that drives disk evolution \citep[e.g.,][]{tabone22a,tabone22b}. 
However, in past theoretical studies, only an isolated protoplanetary disk was considered \citep[see review by][]{pascucci23}. 
Since the magnetic field lines penetrating the disk are connected to the remnant of the star-forming cloud or interstellar medium, the disk wind and magnetic braking should be considered with such a large-scale or cloud core-scale structure (or configuration). 

The purpose of this study is to connect the main accretion phase of star formation to the subsequent phase of protoplanetary disk evolution.
Stars form in gravitationally collapsing cloud cores \citep{larson69,masunaga00}, and 
a rotationally supported disk forms following protostar formation \citep{machida11, dapp12}. 
Observations establish that low-velocity outflows and high-velocity jets appear in the early star formation stage, and their driving mechanisms have been unveiled \citep{Lee14, bjerkeli16,hirota17,matsushita17,tabone17,alves17, matsushita21,sato23,harada23}. 
These flows (outflows and jets) are associated with rotating disks around young stars \citep{hirota17,Lee17, matsushita21,lopez23,omura24}. 
Simulations show that the remnant of the cloud core, the so-called infalling envelope, remains around the protostellar system (protostar and disk) after protostar formation \citep{machida13}.  
Thus, both protostar and disk grow with mass supply from the infalling envelope \citep{tsukamoto22}.
In addition, mass ejection occurs from the protostellar system during the main accretion phase \citep{machida12,tomida17,machida19,Basu24}. An excess angular momentum and a significant part of the infalling gas are expelled from the star-forming cloud core by the 
mass ejection, i.e., protostellar outflow and jet \citep{tomisaka98,joos12,matsushita21}.  
The gravitational potential of the cloud core becomes shallow as a whole with time because the strong mass ejection decreases the mass of  the star-forming cloud.  
Then, the envelope is no longer gravitationally bound and begins to dissipate. 
Consequently, the supply of the mass onto the disk is significantly reduced and the main accretion phase ends \citep{nakano95}. 
Here, we follow the evolution to $\sim 1.5\times10^5$\,yr after protostar formation, to study the transition from an early-stage star-disk-outflow system to a later-stage star-disk system with a depleted envelope. We start with a prestellar cloud, hence this is a self-consistent approach to model the disk by including its formation and subsequent interaction with the parent envelope, using a three-dimensional nonideal magnetohydrodynamic (MHD) simulation. 

The structure of this paper is as follows.
Numerical settings and methods are described in \S\ref{sec:settings}, and calculation results are presented in \S\ref{sec:wholeevo}.
We discuss the driving of the disk wind and the transfer of angular momentum and mass in the disk due to the disk wind in \S\ref{sec:discussion}. 
A summary is presented in \S\ref{sec:summary}.

\section{Initial Conditions and Numerical Settings}
\label{sec:settings}
Our simulation is the same as in \citet{tomida17} and \citet{aso20}. 
Since the initial condition and numerical settings were already explained in these studies, we explain them here only briefly. 
We use our nested grid code \citep{machida04, machida05a, machida07, machida10}. 
The basic equations are described in \citet{machida11} and \citet{tomida17}, while the numerical settings are detailed in \citet{machida13}.

As the initial state, the cloud with a critical Bonnor-Ebert density profile is adopted with a number density $n_{c,0}=6\times10^5\,\cc$ and an isothermal temperature $T_0=10$\,K. 
The cloud density is enhanced by a factor of 2 to promote the contraction. 
The initial cloud mass $M_{\rm cl}$ and radius $R_{\rm cl}$ are $M_{\rm cl}=1.25\,\msun$  and $R_{\rm cl}=6.13\times10^3$\,au, respectively. 
A uniform gas with the number density of $n_{\rm ext}=8.5\times10^4\,\cc$ ($\rho_{\rm ext}=3.4\times10^{-19}$\,g\,cm$^{-3}$) is distributed outside the cloud ($r>R_{\rm cl}$).
A uniform magnetic field ($B_0=51\,\upmu$G) is imposed in the whole computational domain, while a rigid rotation ($\Omega_0=2.0\times10^{-13}$\,s$^{-1}$) is adopted within the cloud core. 
The magnetic field is fixed to its initial value on the surface of the coarsest grid (for details, see \S\ref{sec:numericaleffects}). 
The magnetic vectors are set to be aligned with the rotation vector of the initial cloud. 
The mass-to-flux ratio of the cloud core normalized by the critical value $(2\pi G^{1/2})^{-1}$ is $\mu_0=3$. 
The ratio of the thermal $\alpha_0$, rotational $\beta_0$, and magnetic energy $\gamma_0$ to the gravitational energy are $\alpha_0=0.42$, $\beta_0=0.02$, and $\gamma_0=0.1$, respectively. 
The parameters and properties of the initial cloud core are also the same as in \citet{koga22} and \citet{koga23}.

At the beginning, five levels of nested grids are prepared, in which each grid level is denoted by $l$ and each grid is composed of points ($i$, $j$, $k$) = (64, 64, 32).  
The box size and cell width of the first level grid are  $L (1) = 1.96\times10^5$\,au and $h(1)=3.07\times10^3$\,au, respectively. 
The grid size and cell width halve with each increment of the grid level $l$. 
The finest grid level is $l=13$, which has $L(13)=480$\,au and $h(13)=0.749$\,au, respectively. 
The initial cloud are embedded in the fifth level of grid that has a size of $L(5)=1.23\times10^4$\,au, corresponding to the diameter of the initial cloud.  
The grids of $l=1$--4  (i.e., the large computational domain outside the initial cloud) are set so as to suppress the artificial reflection of Alfv\'en and sound waves at the boundary and to trace the propagation of the outflow beyond the scale of the initial cloud  \citep{machida13,machida20}. 
In addition, an artificial boundary is imposed on the surface of $r=R_{\rm cl}$ to suppress unlimited gas influx from the outside the cloud core $r>R_{\rm cl}$.
The gas influx is prohibited on the surface of $r=R_{\rm cl}$ and gravity (gas self-gravity and protostellar gravity) is switched off in the region of $r>R_{\rm cl}$. 
Thus, the ambient medium outside the cloud ($r>R_{\rm cl}$) maintains a quiescent state without any force before the outflow exceeds the boundary of $r=R_{\rm cl}$. 
On the other hand, the gas outflow is permitted on the boundary $r=R_{\rm cl}$ when the radial velocity $v_r$ exceeds the sound speed $c_s$ \citep[see for example,][]{machida09}.  
The mass ejection by the protostellar outflow is adequately treated in this way, as described in \citet{machida13}. 
With this method, the mass of the (initial) cloud is completely conserved in the region of $r<R_{\rm cl}$, including the sink mass (see below), during the epoch before the outflow reaches the boundary of the cloud $r=R_{\rm cl}$, because gas leakage is completely prohibited \citep{machida13}.
With these treatments, we can adequately investigate the dissipation of the star-forming cloud, as already investigated in our previous studies \citep[e.g.,][]{machida12}. 
The necessity and influence of the large region outside the cloud $r > R_{\rm cl}$ are discussed in \S\ref{sec:numericaleffects}.

To realize a long-term calculation, we use a sink method. 
At the center of the finest grid $l=13$, a sink cell of radius $r_{\rm sink}$ is produced when the central density reaches $n_{\rm thr}$.
The sink parameters are $n_{\rm thr}=10^{13}\,\cc$ and $r_{\rm sink}=1$\,au, respectively.
Details of the implementation of our sink method are given in \citet{machida10} and \citet{machida14}. 

Finally, we explain the difference between the present study and  \citet{tomida17} and \citet{aso20}.
In those studies, the simulation data was used for performing synthetic observations.
In \citet{tomida17}, using a snapshot at $4.61 \times10^4$\,yr after protostar formation, a synthetic observation of the spiral structure was produced and then compared with the observation of Elias 2-27.
At that time, the simulation was completed until a time after protostar formation of   $t_{\rm ps} = 4.7\times10^4$\,yr. 
Later, \citet{aso20}, performed synthetic observations using five snapshots at $t_{\rm ps}=4.71\times10^3$, $1.26\times10^4$, $2.2\times10^4$, $3.7\times10^4$ and $7.3\times10^4$\,yr. 
At that time, the simulation was completed until  $\sim7\times10^4$\,yr.
The simulation used in this paper started in 2016 and it has now reached $1.5\times10^5$\,yr after protostar formation. 
Although the simulation is still running now, the time step is small and it is difficult to further extend the integration time. 
Thus, we summarize the result up until $1.5\times 10^5$\,yr after protostar formation. 
Since this is not synthetic observation study, we leave a presentation of synthetic observations in the late phase to a subsequent paper.

\section{Results}
\label{sec:wholeevo}
\begin{figure*}
\begin{center}
\includegraphics[width=0.9\columnwidth]{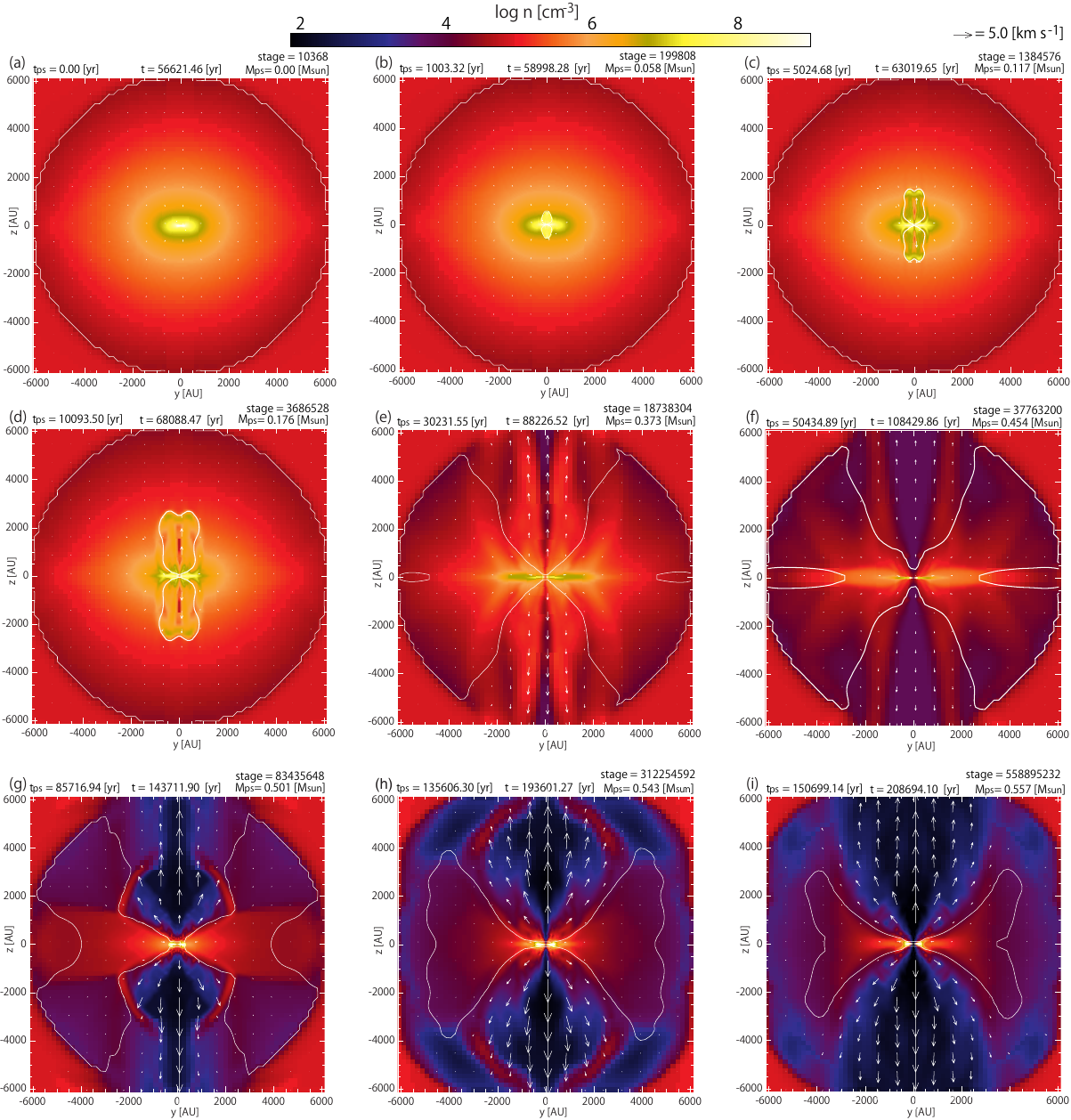}
\end{center}
\caption{
Density (color) and velocity (arrows) distributions at different epochs on the $x=0$ plane. 
The elapsed time $t_{\rm ps}$ after protostar formation and the time $t$ after the beginning of the cloud collapse are labeled in the upper part of each panel. 
The protostellar mass $M_{\rm ps}$ is also labeled in each panel. 
The white line corresponds to the boundary between the infalling ($v_r<0$) and outflowing ($v_r>0$) regions within which the gas falls toward the center. 
An animated version of this figure is available.
In the animation, the times sequence of density and velocity distribution as in Fig.~\ref{fig:1} until the end of the simulation from the beginning is shown. 
The The duration of the animation is 219 seconds.
(High-resolution animation see 
https://archive.iii.kyushu-u.ac.jp/public/G6o0gpIIqCsu5sISEeKLq0Qk7eJ7LkTK7CE-kzMpOgtA)
}
\label{fig:1}
\end{figure*}

Figure~\ref{fig:1} shows the time sequence of the density and velocity distributions for $\sim1.5\times10^5$\,yr after protostar formation. 
Figure~\ref{fig:1}{\it a} shows the cloud structure at the protostar formation epoch. 
At this epoch, the central region has an elliptical structure. 
The evolution of the outflow within the cloud can be seen in Figures~\ref{fig:1}{\it b}--{\it d}.
The outflow has a size of $\sim5000$\,au at $10^4$\,yr after protostar formation (Fig.\ref{fig:1}{\it d}).
The outflow reaches the cloud boundary corresponding to the white incircle in Figure~\ref{fig:1}{\it a}--{\it d}. 
Then, the outflow widens and ejects a large fraction of the cloud mass  (Fig.~\ref{fig:1}{\it e}).  
In Figure~\ref{fig:1}{\it f},  the region inside the white line corresponds to the infalling envelope where the gas falls toward the center. 
Figures~\ref{fig:1}{\it f} and {\it g} indicate that the gas near the equatorial plane in the range of $\vert y \vert \gtrsim 4000$\,au does not fall toward the center and is outflowing. 
The equatorial outflow is explained below. 
The outflow, which is driven by the first core before protostar formation, weakens after the mass ejection rate peaks at $t_{\rm ps}\sim3.0\times10^4$\,yr (see below).
Comparing Figure~\ref{fig:1}{\it e} and Figure~\ref{fig:1}{\it f}, the density and speed of the outflow seen in Figure~\ref{fig:1}{\it f} are smaller than those of the outflow in Figure~\ref{fig:1}{\it e}.

Figure~\ref{fig:1}{\it g} shows that a different mode of the outflow emerges around the center.
The outflow that appears later has a high speed and creates a cavity or shell-like structure within the outflow that appeared previously, as seen in Figure~\ref{fig:1}{\it h}. 
Hereafter, we call the later outflow the disk wind and the former the outflow.  
Although the ejected mass is smaller in the disk wind than in the outflow, the disk wind steadily 
reduces the size of 
the infalling envelope, as 
seen in Figures~\ref{fig:1}{\it h} and \ref{fig:1}{\it i}.

Figure~\ref{fig:1} also shows that the density of the infalling envelope becomes gradually lower as time progresses. 
This is because of the finite mass available in the cloud ($r < R_{\rm cl}$).
The number density of the infalling envelope around $r=R_{\rm cl}$ is $\sim10^5\,\cc$ at the protostar formation epoch, while it is $\sim10^2\,\cc$ at the end of the simulation. 
As seen in Figures~\ref{fig:1}{\it h} and \ref{fig:1}{\it i}, only a butterfly-shaped infalling envelope with a low density remains around the protostellar system at the end of the simulation. 
The infalling envelope is almost dissipated by the end of the simulation.

\begin{figure*}
\begin{center}
\includegraphics[width=0.9\columnwidth]{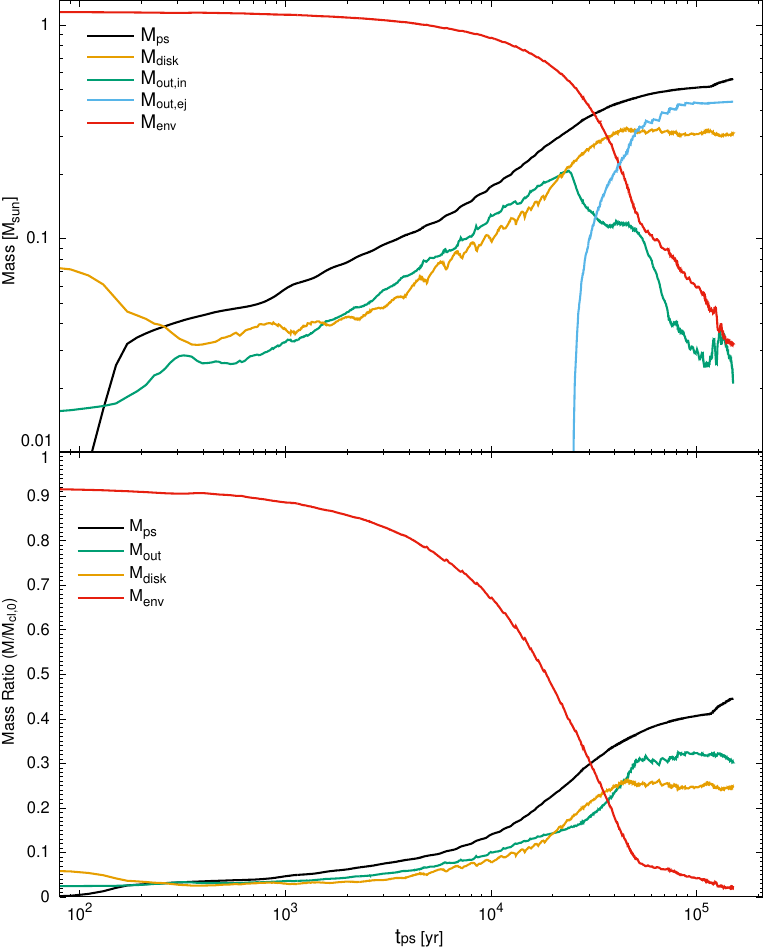}
\end{center}
\caption{Mass and mass fractions versus $t_{\rm ps}$, the elapsed time after protostar formation.
Top: Mass of the protostar (black), circumstellar disk (orange), outflow within the core (green), ejected outflow (blue), and infalling envelope (red).
Bottom: Mass of the protostar (black), circumstellar mass (orange), outflow ($M_{\rm out}= M_{\rm out,in}+M_{\rm out,ej}$, green), and infalling envelope (red), all normalized by the initial cloud mass.
}
\label{fig:2}
\end{figure*}

Figure~\ref{fig:2} shows the mass (top) and the mass fraction (bottom) of the protostar, circumstellar disk, outflow, and infalling envelope, respectively, against the elapsed time $t_{\rm ps}$ after protostar formation. 
The protostellar mass $M_{\rm ps}$ is defined as the mass that has fallen into the sink. 
The disk is defined as the region where the gas exceeds the disk density $n_{\rm d}$ (or $\rho_{\rm d}$).
Here, $n_{\rm d}$ is the minimum value that fulfills the following criteria:  
(1) the rotation velocity dominates the radial velocity ($v_\phi > 2 \vert v_r \vert $); 
(2) the radial velocity is larger than 80\% of the Keplerian velocity ($v_\phi > 0.8\, v_{\rm kep}$ where $v_{\rm kep}$ is the Keplerian velocity); and 
(3) the radial velocity is much smaller than the sound velocity ($v_r<0.1 c_s$ where $c_s$ is the local sound velocity). 
The third criterion is necessary so as to exclude the outflow. 
The disk mass $M_{\rm disk}$ is estimated as the mass of the region where $n>n_{\rm d}\ ({\rm or} \ \rho>\rho_{\rm d})$.  
The outflow is defined as the region where the radial velocity is greater than 10\% of the local sound speed ($v_r>0.1\,c_{\rm s}$). 
The outflow mass is estimated by integrating the outflow region having $v_r > 0.1\, c_{\rm s}$. 
We distinguish between the outflow mass $M_{\rm out,in}$ within the cloud core ($r<R_{\rm cl}$) and the outflow mass $M_{\rm out, ej}$ outside the cloud core ($r>R_{\rm cl}$). 
Within the cloud ($r<R_{\rm cl}$), we define  the envelope as the region not belonging to either the disk or the outflow. 
Thus, the envelope mass is estimated as $M_{\rm env}=M_{\rm tot}-M_{\rm disk}-M_{\rm out, in}$, where $M_{\rm tot}$ is the total mass within the cloud. 
The procedure to determine each mass adopted in this study is the same as in \citet{machida13} and \citet{tomida17}.

Figure~\ref{fig:2} (top panel) shows that the envelope mass continues to decrease and reaches $M_{\rm env}\simeq 0.03\,\msun$ at the end of the simulation, at a time $t_{\rm ps}=1.51\times10^5$\,yr after protostar formation.  
Since the initial cloud mass is $M_{\rm cl}=1.25\,\msun$, only 3\% of the initial cloud mass remains as the infalling envelope (Fig.~\ref{fig:2} bottom).
Thus, the mass accretion from the infalling envelope to the circumstellar disk almost ends and the system is considered to be in the Class II stage.

The outflow mass $M_{\rm out, in}$ inside the cloud continues to increase for $t_{\rm ps} \lesssim 2\times 10^4$\,yr and decreases for $t_{\rm ps} > 2\times 10^4$\,yr. 
The outflow reaches the cloud boundary at $t_{\rm ps}\sim2\times10^4$\,yr and a large fraction of the mass is ejected from the cloud.  
The ejected outflow mass $M_{\rm out, ej}$ rapidly increases at $t_{\rm ps}\simeq 2.3\times10^4$\,yr and reaches $M_{\rm out, ej}=0.45\,\msun$.
Note that since the ejected outflow mass is estimated as the region having $v_r>0.1\,c_s$ outside the cloud, a small amount of the ambient mass ($\lesssim0.05\msun$) distributed outside of the cloud is included. 
As a result, about 30\% of the initial cloud mass is ejected from the cloud by the end of the simulation. 

\begin{figure*}
\begin{center}
\includegraphics[width=0.9\columnwidth]{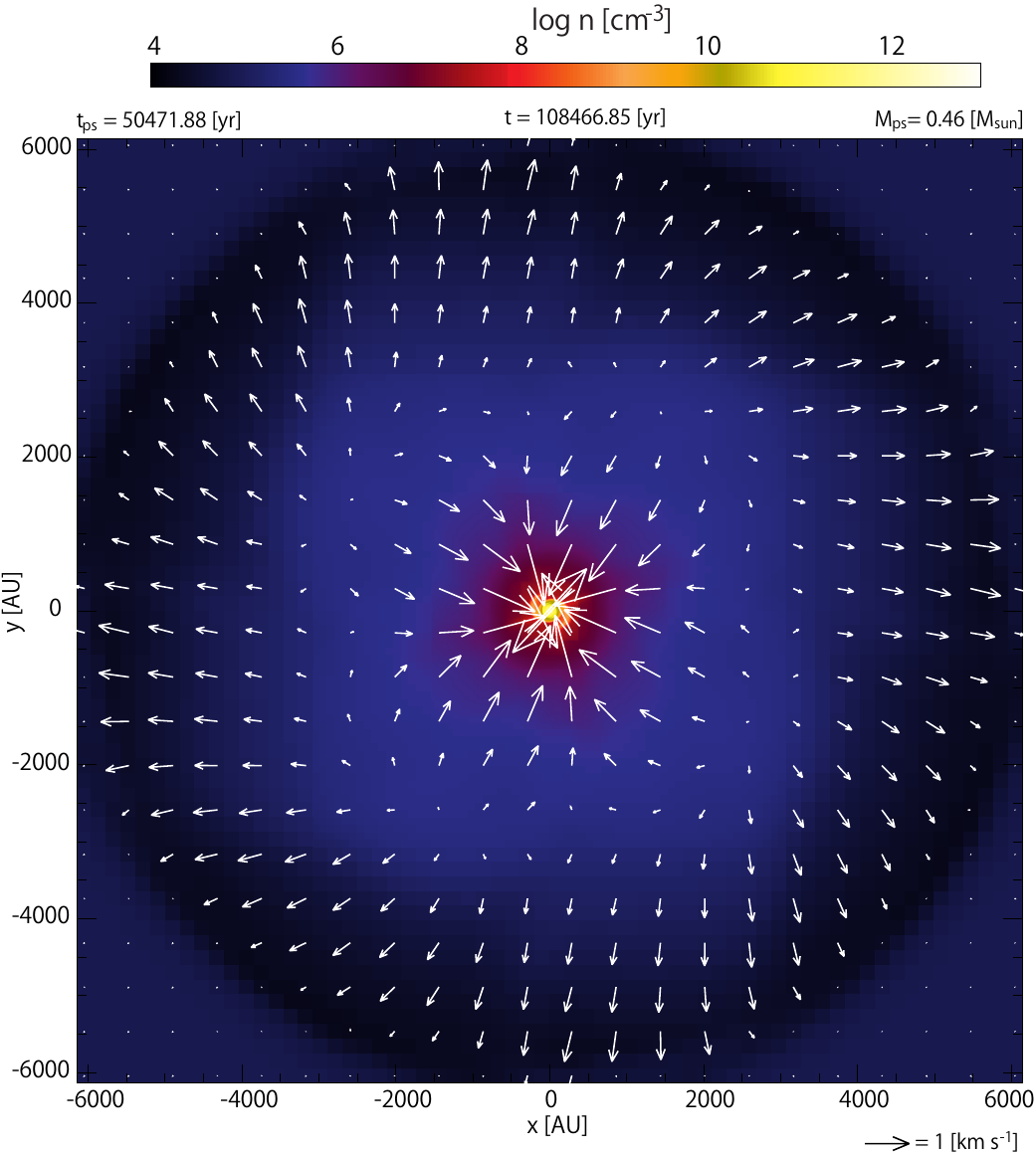}
\end{center}
\caption{
Density (color) and velocity (arrows) distributions on the $z=0$ plane. 
The elapsed time $t_{\rm ps}$ after protostar formation, the time $t$ after the beginning of the cloud collapse, and protostellar mass $M_{\rm ps}$ are labeled.
}
\label{fig:3}
\end{figure*}

\begin{figure*}
\begin{center}
\includegraphics[width=0.9\columnwidth]{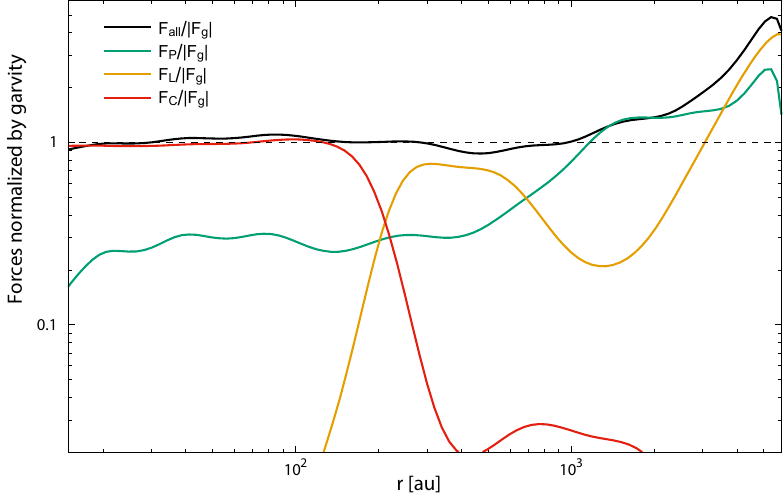}
\end{center}
\caption{
The ratios of all ($F_{\rm all}$), gas pressure gradient ($F_{\rm P}$), Lorentz ($F_{\rm L}$), and centrifugal ($F_{\rm c}$) forces to gravity ($F_{\rm g}$) on the equatorial plane are plotted against radius at the same epoch as Fig.~\ref{fig:3}, in which each force is azimuthally averaged. 
$F_{\rm all}$ is the sum of all forces except for gravity $F_{\rm all}=F_{\rm P} +F_{\rm L} +F_{\rm c}$.
The dashed line indicates where each force is equal to gravity. 
}
\label{fig:3b}
\end{figure*}

After the head of the outflow reaches the boundary between the (star-forming) cloud ($r<R_{\rm cl}$) and interstellar medium ($r>R_{\rm cl}$), the cloud begins to dissipate as a whole, as shown in Figure~\ref{fig:1}{\it e}-{\it i}. 
The outflowing region near and on the equatorial plane can be seen in Figures~\ref{fig:1}{\it e}, {\it f} and {\it g}. 
Figure~\ref{fig:3} shows the density and velocity distributions at $t_{\rm ps}=5.05\times10^4$\,yr, at which time a large fraction of the cloud gas is already ejected from the cloud. 
The figure indicates that the gas is outflowing in the range of $r\gtrsim 2000$\,au on the equatorial plane. 
This outflow in the equatorial plane is a major finding of this simulation; it is not driven by the ram pressure of the disk wind. The disk wind itself does not broaden enough to reach the equatorial plane. 
The pressure gradient and Lorentz force contribute to the emergence of the equatorial outflow as described below. 
As seen in Figure~\ref{fig:1}{\it e}--{\it i}, the gas density (or gas pressure) is high near the equatorial plane. 
For $t_{\rm ps}\gtrsim 2\times10^4$\,yr, a large amount of the mass is ejected by the outflow, indicating that the gravitational potential becomes shallower in this late stage compared to the early stage. 
Thus, the pressure gradient force overcomes the gravity in the region far from the protostar and  the gas distributed in such a region is outflowing and ejected from the cloud. 
This phenomenon was anticipated earlier by \citet{nakano95} in their semi-analytic model. 

The ratio of each force to gravity on the equatorial plane is plotted against the radius in Figure~\ref{fig:3b}.
The figure shows the radial component of the pressure gradient $F_{\rm p}$, Lorentz $F_{\rm L}$ and centrifugal $F_{\rm C}$ forces divided by the absolute value of the radial component of gravity.
Note that we confirmed that the forces, except for gravity, have a positive sign in the radial direction. 
The figure indicates that the centrifugal force is balanced with gravity within $\lesssim200$\,au. 
On the other hand, the pressure gradient force dominates gravity in the range of $\gtrsim1000$\,au.
In addition, the Lorentz force also exceeds gravity in the range of $\gtrsim3000$\,au. 
Thus, both the pressure gradient and Lorentz force contribute to the equatorial outflow shown in Figure~\ref{fig:3}.

Figure~\ref{fig:1}{\it g}-{\it h} shows that as
the cloud mass decreases further, the whole outer region of the cloud begins to dissipate.
The rate of decrease of the envelope declines due to the deficit of the remaining infalling gas. Figure~\ref{fig:2} shows that the envelope mass decreases to about 10\,\% of the initial cloud mass at these epochs (Fig.~\ref{fig:1}{\it g}-{\it h}).

As seen in Figure~\ref{fig:2}, the disk mass continues to increase until $t_{\rm ps} \simeq 4\times10^4$\,yr, during which time the envelope mass dominates the disk mass.
For $t_{\rm ps}\gtrsim 4\times10^4$\,yr, the disk mass dominates the envelope mass and keeps a constant value of $M_{\rm disk}\sim0.3\,\msun$ until the end of the simulation. 
At the end of the simulation, the mass ratio of the protostar, disk, outflow, and envelope to the initial cloud mass are $M_{\rm ps}/M_{\rm cl}=0.44$,  $M_{\rm disk}/M_{\rm cl}=0.24$, $M_{\rm out}/M_{\rm cl}=0.30$ and $M_{\rm env}/M_{\rm cl}=0.02$, respectively.

\begin{figure*}
\begin{center}
\includegraphics[width=0.9\columnwidth]{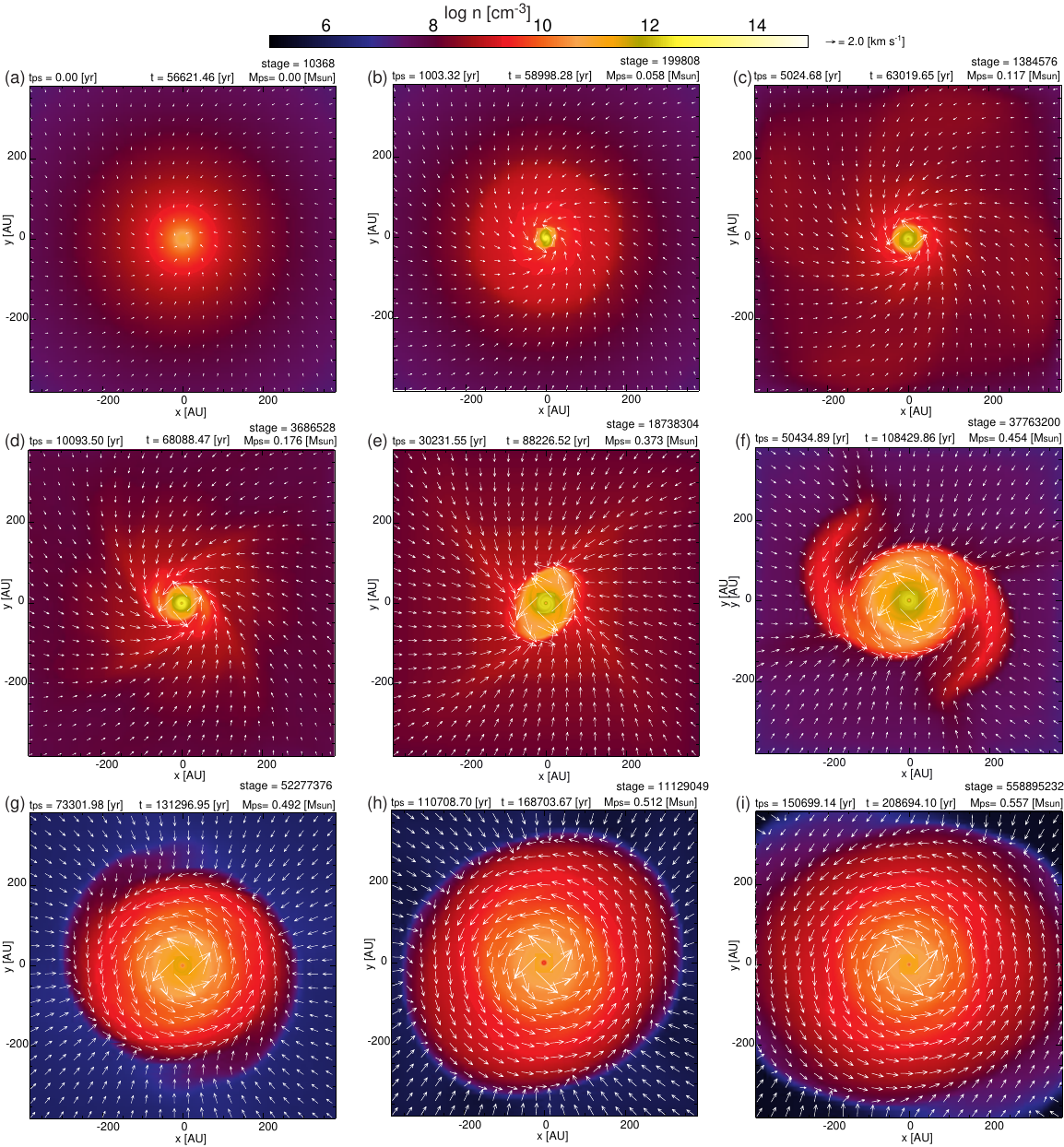}
\end{center}
\caption{
Density (color) and velocity (arrows) distributions at different epochs on the $z=0$ plane. 
The elapsed time $t_{\rm ps}$ after protostar formation and time $t$ after the beginning of the cloud collapse are described in the upper part of each panel. 
The protostellar mass $M_{\rm ps}$ is also described in each panel. 
}
\label{fig:4}
\end{figure*}

The disk evolution for $\sim1.5\times10^5$\,yr after protostar formation is shown in Figure~\ref{fig:4}, in which the density and velocity distributions on the equatorial plane are plotted in each panel. 
As shown in Figure~\ref{fig:4}{\it b}--{\it e}, a tiny disk appears following protostar formation and its size gradually increases with time. 
Recent observations \citep{ohashi23} reveal similar small disks with a lack of substructure during the earliest phases of protostellar evolution. 

A spiral structure caused by gravitational instability can be seen  in Figure~\ref{fig:4}{\it f}.
In Figure~\ref{fig:4}{\it g}--{\it i}, although an apparent spiral cannot be confirmed, the disk continues to increase its size until the end of the simulation. 

\begin{figure*}
\begin{center}
\includegraphics[width=0.9\columnwidth]{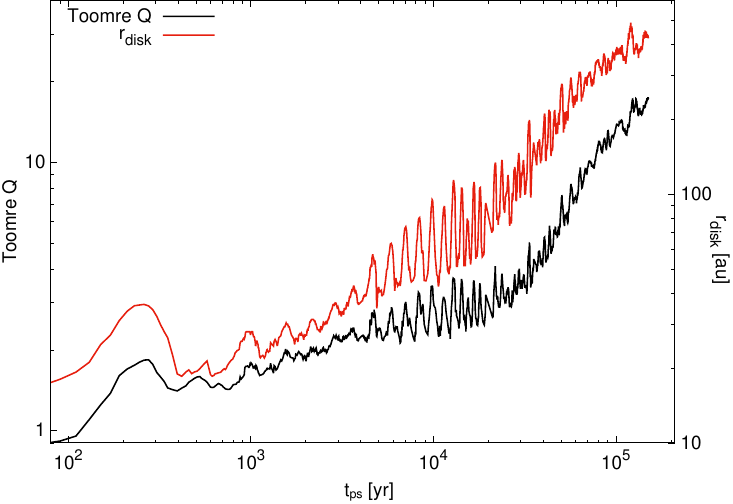}
\end{center}
\caption{
Toomre $Q$ parameter (black; left axis) and disk radius $r_{\rm disk}$ (red; right axis) versus $t_{\rm ps}$, the elapsed time after protostar formation. 
}
\label{fig:5}
\end{figure*}

We estimate the Toomre $Q$ parameter of the disk as
\begin{equation}
Q = \frac{\int_{\rho > \rho_{\rm d}} \frac{c_s \Omega_{\rm Kep}}{\pi G \Sigma}\, \Sigma\, dS }{\int_{\rho > \rho_{\rm d}}\, \Sigma\, dS}\, ,
\end{equation} 
where $c_s$ is the local sound speed, $\Omega_{\rm Kep} \equiv (GM_{\rm ps}/r^3)^{1/2}$ is the Keplerian angular velocity, and
\begin{equation}
\Sigma = \int_{\rho > \rho_{\rm d}} \rho\, dz
\end{equation}
is the disk surface density.
Thus, the $Q$ parameter is weighted by the disk surface density, and represents the degree of the gravitational instability of the whole disk. This is the same method used by \citet{tomida17}.
Figure~\ref{fig:5} shows the Toomre $Q$ parameter and disk radius plotted against the elapsed time after protostar formation.
As time proceeds, both the $Q$ parameter and the disk radius increase while also oscillating.
The oscillation is caused by the gravitational instability of the disk. 
The accretion from the infalling envelope increases the disk surface density and the $Q$ parameter decreases. 
Then, the nonaxisymmetric (or spiral) structure develops and transports the disk angular momentum outward due to gravitational torques.
Therefore, part of the disk mass rapidly falls onto the central protostar, and the disk mass decreases and the $Q$ parameter increases.  
Then, the $Q$ parameter begins to decrease because the disk mass begins to increase due to the accretion from the infalling envelope. 
This process repeats and causes episodic accretion in a manner first demonstrated by \cite{vorobyov05,vorobyov06}.
The amplitude of the oscillation in $Q$ decreases as the disk size increases. 
The disk has a size of $400-500$\,au for  $t_{\rm ps}>10^5$\,yr. 
Such large-sized Keplerian disks have been confirmed in recent observations \citep[e.g.,][]{hara13,yen14,huang21}. 
As seen in Figures~\ref{fig:4}{\it h} and {\it i}, the disk maintains a nearly axisymmetric structure without apparent spirals. 
Thus, the disk becomes stable with time. 

\begin{figure*}
\begin{center}
\includegraphics[width=0.9\columnwidth]{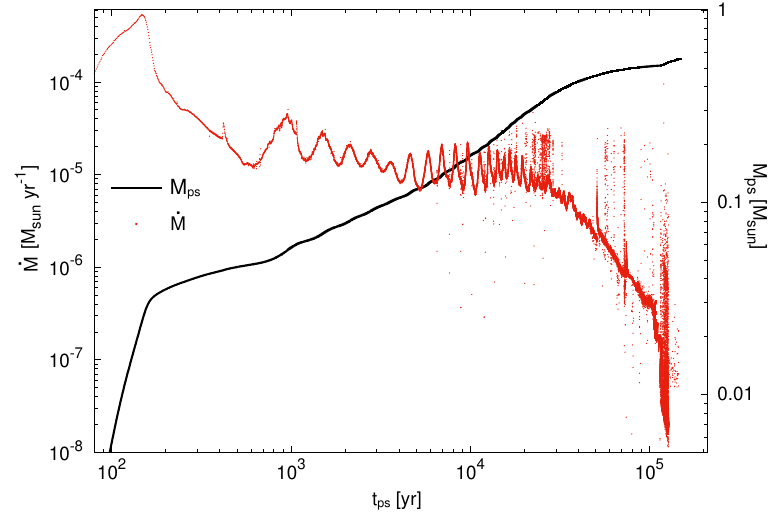}
\end{center}
\caption{
Mass accretion rate $\dot{M}$ (red dots; left axis) and protostellar mass $M_{\rm ps}$ (solid black line; right axis) against the time after protostar formation $t_{\rm ps}$.
}
\label{fig:6}
\end{figure*}

The mass accretion rate onto the protostar (or sink) is plotted against the elapsed time by the red dots in Figure~\ref{fig:6}. 
For $t_{\rm ps}\lesssim 3\times 10^4$\,yr, although the mass accretion rate strongly oscillates, it roughly maintains a value $\dot{M}\sim 10^{-5}\,\msun$\,yr$^{-1}$. 
Comparison between Figure~\ref{fig:5}  and Figure~\ref{fig:6} indicates that the oscillation in the mass accretion rate is well synchronized with that of the $Q$ parameter or disk radius. 
This confirms that the episodic accretion is caused by the disk instability.
The mass accretion rate begins to decrease at $t_{\rm ps}\sim  3\times10^4$\,yr, at which time the envelope begins to dissipate (see Fig.~\ref{fig:1}{\it e}). 

For $t_{\rm ps}\gtrsim 3\times 10^4$\,yr, the mass accretion rate decreases by about three orders of magnitude (from $\sim10^{-5}$ to $10^{-8}\,\msun$\,yr$^{-1}$). 
However, even for $t_{\rm ps}\gtrsim 3\times10^4$\,yr, the mass accretion rate is transiently enhanced because the gravitational instability sometimes occurs and an accretion burst happens. 
The mass accretion rate in the accretion burst phase is 1--3 orders of magnitude greater than that in the quiescent phase.
This indicates that the accretion luminosity can be enhanced by 1--3 orders of magnitude in the accretion burst phase. 
However, the cycle of the oscillation in the mass accretion  (or the cycle between the burst and  quiescent phases) for $t_{\rm ps}\gtrsim 3\times10^4$\,yr is much longer than that for $t_{\rm ps}\lesssim  3\times10^4$\,yr, because the mass accretion from the infalling envelope to the disk significantly decreases in the late phase. 
Reflecting the history of the mass accretion rate  (red dots in Fig.~\ref{fig:6}), the increase rate of the protostellar mass (black solid line in Fig.~\ref{fig:6}) is greater for $t_{\rm ps}\lesssim 3\times10^4$\,yr than for $t_{\rm ps}\gtrsim  3\times10^4$\,yr. 

\section{Discussion}
\label{sec:discussion}

\subsection{Outflow Driving}
\label{sec:outflowdriving}
\begin{figure*}
\begin{center}
\includegraphics[width=0.9\columnwidth]{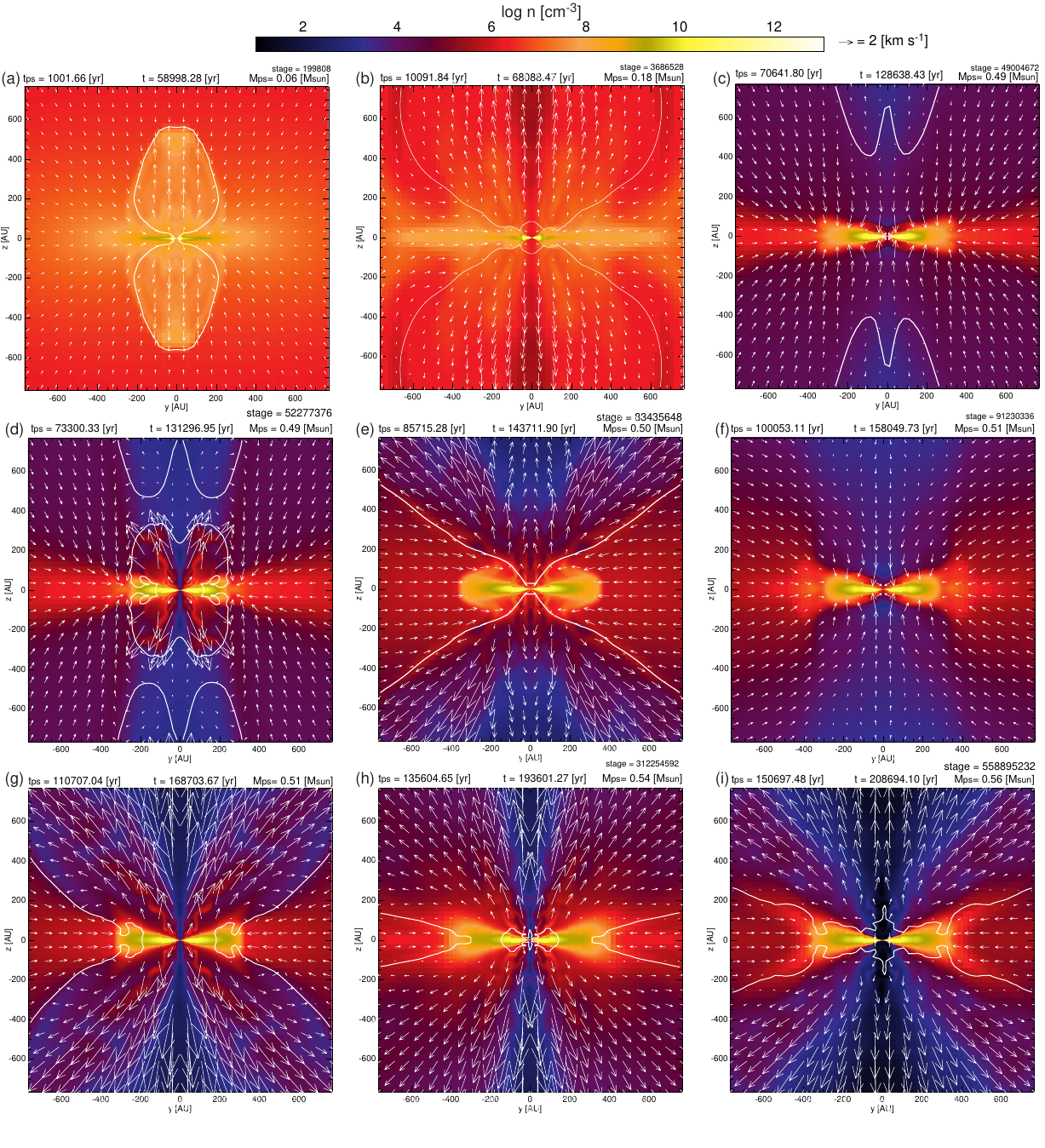}
\end{center}
\caption{
Same as in Fig.~\ref{fig:1}, but different spatial scale. 
}
\label{fig:7}
\end{figure*}
In this subsection, we discuss why the outflow is activated again in a later evolutionary stage. 
The evolution of the outflow is shown in Figure~\ref{fig:7}, in which the density and velocity distributions on the $x=0$ plane are plotted in each panel. 
As described in many previous studies \citep{tomisaka02,machida04,banerjee06,tomida13,vaytet18}, in the star-forming cloud, the outflow is driven by the first core before protostar formation \citep{tomisaka98}. 
The first core becomes the rotationally supported disk following protostar formation, and the outflow originated from the first core continues to be driven after protostar formation \citep{machida11}.  
As seen in Figure~\ref{fig:7}{\it a}, the outflow has a size
$\sim 500$\,au
even when the protostellar mass is as small as $M_{\rm ps}=0.06\,\msun$, because the outflow is initiated before protostar formation \citep{machida06,hennebelle08}.
The outflow evolves in size in the early accretion phase. 
In addition, a strong mass ejection from near the disk can be confirmed in Figure~\ref{fig:7}{\it b}. 
The outflow at the early phase is driven from the pseudodisk, with the strongest flow coming from the interface of the disk and pseudodisk \citep{Basu24}.
The outflow temporarily stops (Fig.~\ref{fig:7}{\it c}), before the mass ejection  starts again (Fig.~\ref{fig:7}{\it d} and {\it e}). 
The mass ejection temporarily stops again at $t_{\rm ps}\simeq 1.0\times10^5$\,yr (Fig.~\ref{fig:7}{\it f}), however a mass ejection from the disk surface subsequently begins (Fig.~\ref{fig:7}{\it g}).  
The disk surface and infalling envelope are ablated by the ejected gas. 
We call this mass ejection in the later phase the disk wind, as described above.
The disk wind lasts and does not significantly weaken by the end of the simulation (Fig.~\ref{fig:7}{\it i}). 

\begin{figure*}
\begin{center}
\includegraphics[width=0.9\columnwidth]{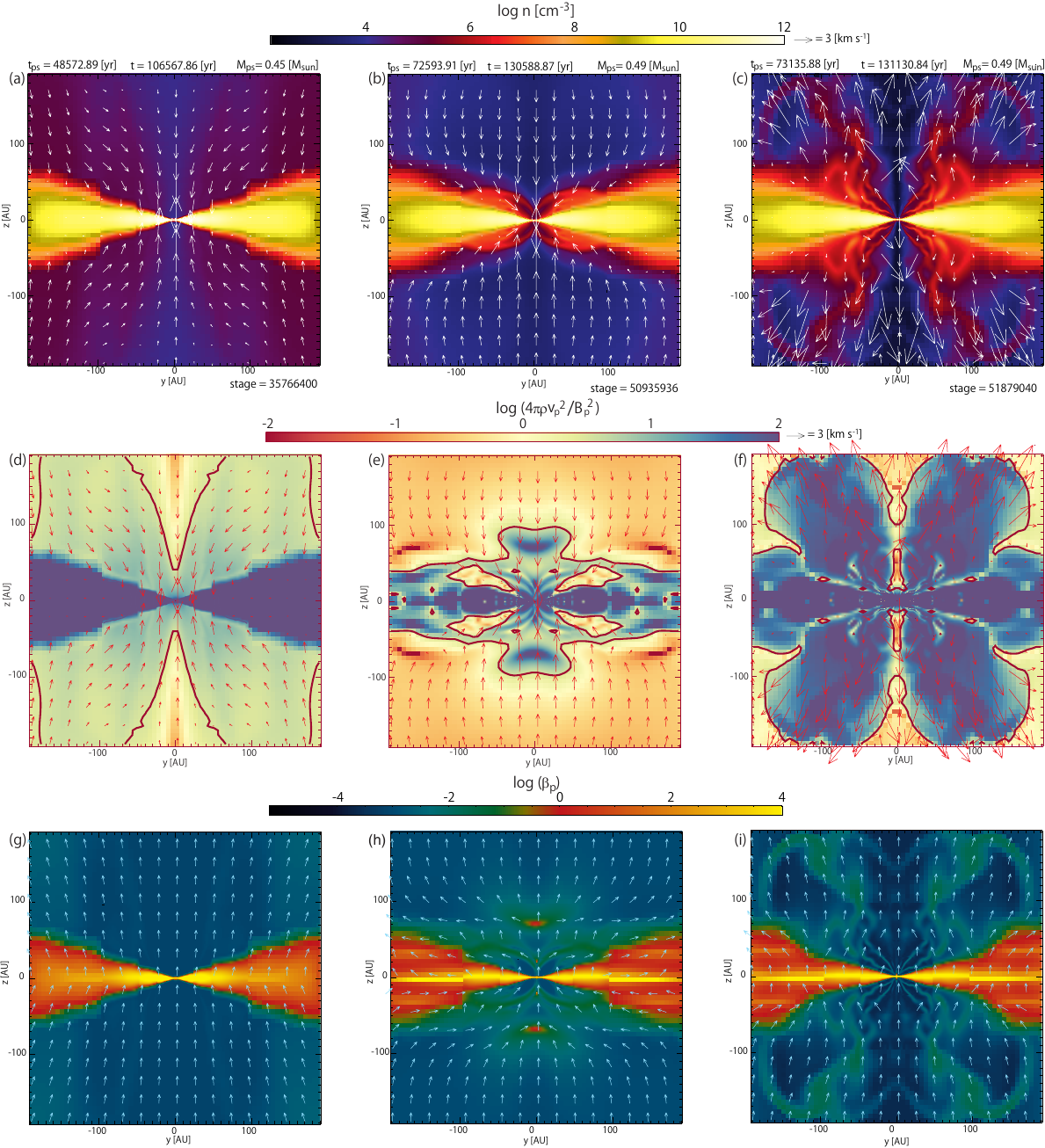}
\end{center}
\caption{
Top panels: Density (color) and velocity (arrows) distributions at different epochs on the $x=0$ plane. 
The elapsed time $t_{\rm ps}$ after protostar formation and time $t$ after the beginning of the cloud collapse are labeled in the upper part of each panel. 
The protostellar mass $M_{\rm ps}$ is also described in each panel. 
Middle panels: Ram pressure divided by square of Alfv\'en velocity (color) and gas velocity (arrows) on the $x=0$ plane at the same epochs as in the top panels. 
Red contour corresponds to $4\pi \rho v_{\rm p}^2/B_{\rm p}^2=1$, within which the ram pressure dominates the square of Alfv\'en velocity.
Bottom panels: Plasma beta ($\beta_p \equiv (8 \pi P)/B^2$) and magnetic field vectors ($\vect{B}/\vert \vect{B} \vert$) on the $x=0$ plane at the same epochs as in the top panels. 
}
\label{fig:8}
\end{figure*}

In our previous studies \citep{machida11,machida20}, we showed that the outflow gradually weakens as the envelope dissipates. 
In this study, the outflow weakens as the infalling envelope dissipates, as in previous studies. 
However, the outflow (or disk wind) is activated again in the further long-term time integration. 
To investigate the outflow driving condition, the density (top panels), ram pressure divided by the square of the Alfv\'en velocity ($4\pi \rho v_{\rm p}^2/B_{\rm p}^2$) and plasma beta ($\beta_p \equiv (8 \pi P)/B^2$) on the $x=0$ plane  before and after the reactivation of the outflow are plotted in Figure~\ref{fig:8}, where $v_{\rm p}$ and $B_{\rm p}$ are the poloidal components of velocity and magnetic field, respectively. 
As seen in Figure~\ref{fig:8}{\it a} and {\it b}, the density of the infalling envelope gradually lowers. 
The density of the infalling envelope at $t_{\rm ps}=4.9\times10^4$\,yr is $n\sim10^5-10^6\,\cc$ (Fig.~\ref{fig:8}{\it a}), while that at $t_{\rm ps}=7.3\times10^4$\,yr is $n\lesssim 10^4\,\cc$  (Fig.~\ref{fig:8}{\it b}). 

The ratio of the ram pressure to magnetic pressure is plotted in the middle panels of Figure~\ref{fig:8}.
Figure~\ref{fig:8}{\it d} indicates that the ram pressure dominates the magnetic pressure in the region above and below the circumstellar disk.
Thus, it is difficult to drive the outflow from the disk at this epoch, because the ram pressure suppresses the outflow driving \citep{matsushita17,machida20,machida21}. 
Note that the ram pressure at this epoch is less than in earlier epochs, when the envelope mass was greater. However, in those epochs the infall along primarily equatorial regions was strong enough to power an outflow at wide angles above and below the disk-pseudodisk region \citep[for details, see][]{Basu24}, overcoming the ram pressure from infall along those directions. 
As the ram pressure-dominated region shrinks, the magnetic pressure becomes dominant around the disk at $t_{\rm ps}=7.3\times10^4$\,yr (Fig.~\ref{fig:8}{\it e}). 
Then, a strong mass ejection occurs again (Figs.~\ref{fig:8}{\it c} and {\it f}). 
As seen in the bottom panels of Figure~\ref{fig:8}, the plasma beta is much smaller than unity ($\beta_{\rm p} \ll 1$) at any time in the region above and below the disk. 
In addition, the comparison between the middle and bottom panels means that the ram pressure always dominates the gas pressure in the region above and below the disk. 

As shown in Figures~\ref{fig:1} and \ref{fig:7}, the outflow (or disk wind) is reactivated after the outflow driven before protostar formation significantly weakens (see Fig.~\ref{fig:7}{\it c} and {\it d}). 
The ram and magnetic pressures should determine whether or not the outflow can be driven from the disk \citep{blandford82,uchida85,machida20,machida21}, and both pressures vary over time. 
The ram pressure $\rho v_{\rm p}^2$ increases as the infall velocity $v_{\rm p}$ increases, while it decreases as the density of the infalling envelope $\rho$ decreases. 
The infall velocity increases with time, because it roughly corresponds to the free-fall velocity $(2GM_{\rm ps}/r)^{1/2}$ and the protostellar mass $M_{\rm ps}$ gradually increases. 
Thus, the ram pressure increases with time if only considering the free-fall velocity.
On the other hand, as shown in Figure~\ref{fig:1}, the density of the infalling envelope decreases with time, resulting in the decrease of the ram pressure. 
Therefore, the time variation of the ram pressure $\rho v_{\rm p}^2$ is not simple, since the (free-fall) velocity ($v_{\rm p}$) increases and the density of the infalling envelope $\rho$ decreases during the star formation process. 

During the main accretion phase, the protostellar mass continues to increase because a significant amount of the mass remains in the infalling envelope (Fig.~\ref{fig:2}). 
Then, the protostellar mass (or free-fall velocity) does not significantly increase after the protostellar mass dominates the mass of the infalling envelope. 
After that, the density of the infalling envelope decreases as the infalling envelope dissipates (Fig.~\ref{fig:1}), leading to a decrease of the ram pressure.  
Thus, the ram pressure increases during the main accretion phase, while it decreases after the dissipation of the infalling envelope becomes significant.
Although the driving condition of the outflow (or disk wind) depends also on the magnetic field strength of the disk, the middle and bottom panels of Figure~\ref{fig:8} indicate that the magnetic pressure overwhelms both the thermal and ram pressures after the protostellar mass reaches $\sim0.5\,\msun$, at which time the envelope mass is $\ll 0.1\,\msun$ (Fig.~\ref{fig:2}). 

In summary, the disk wind is naturally driven by the circumstellar disk with a significant decrease of the density of infalling envelope after the main accretion phase. 
Although the disk wind is temporarily inactive in a short time after the protostellar mass dominates the mass of the infalling envelope, it appears again and continues to be driven from the disk by the end of the simulation. 
After the main accretion phase ends, the protostellar mass does not significantly increase but the density of the infalling envelope continues to decrease.
Thus, the ram pressure continues to decrease.
Therefore, it is expected that the disk wind, which is driven by the magnetic effects, continues to be driven by the disk as long as the circumstellar disk exists, because the ram pressure cannot suppress the wind driving.

\subsection{Angular Momentum Transfer by Magnetic Effects}
\begin{figure*}
\begin{center}
\includegraphics[width=0.9\columnwidth]{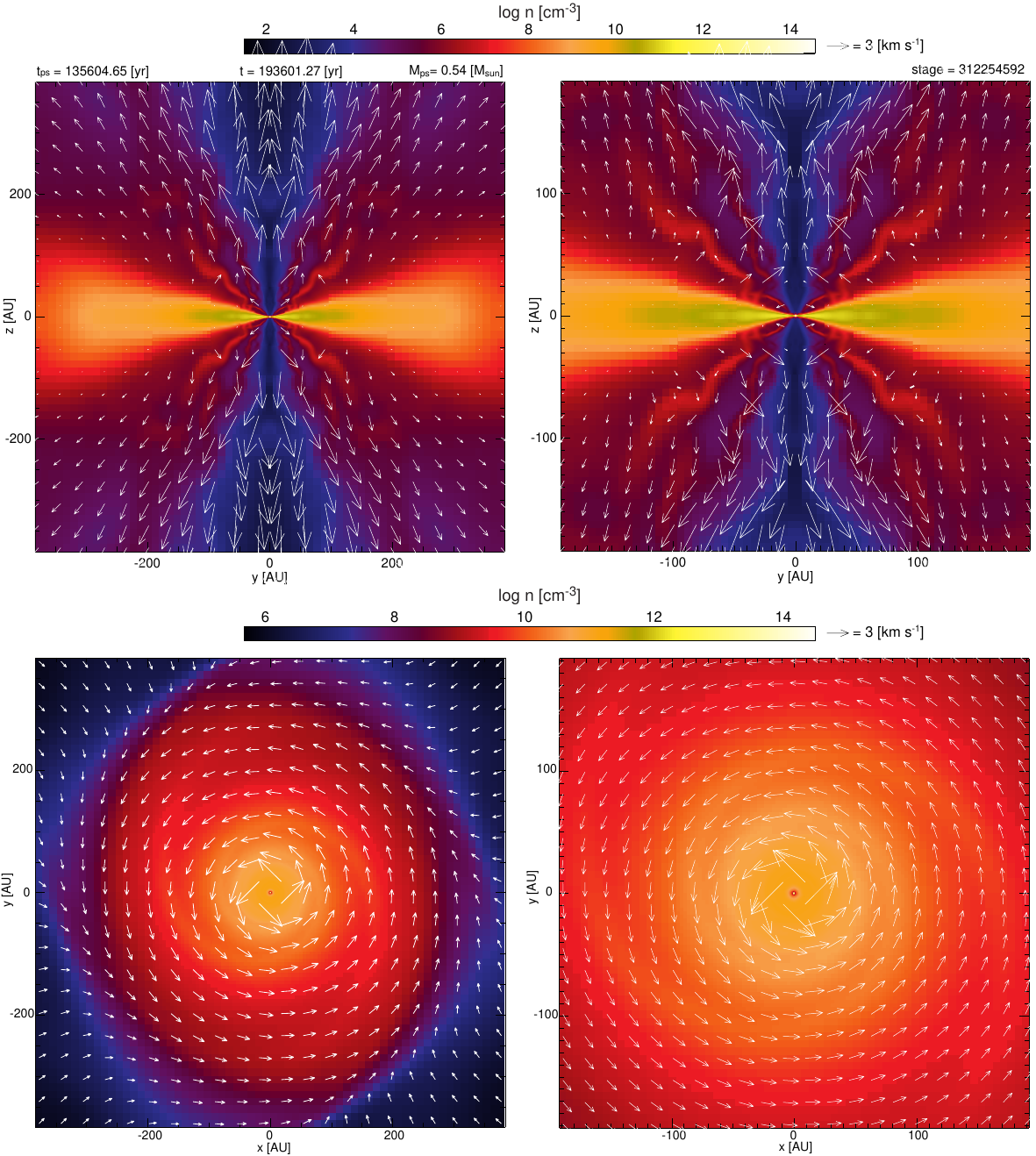}
\end{center}
\caption{
Density (color) and velocity (arrows) distributions on the $x=0$ (top) and $z=0$ (bottom) plane. 
The elapsed time $t_{\rm ps}$ after protostar formation and the time $t$ after the beginning of the cloud collapse are labeled in the top left panel. 
The protostellar mass $M_{\rm ps}$ is also labeled.
Each right panel is an enlarged view of the central region of each left panel.  
}
\label{fig:9}
\end{figure*}

In this subsection, we focus on the angular momentum transport within the disk by magnetic effects.
Figure~\ref{fig:9} shows the density and velocity distributions at almost the end of the simulation. 
The figure indicates that the wind is driven directly from the rotating disk and the flow speed gradually increases as the distance from the disk increases. 
Although we stopped the simulation at $1.5\times10^5$\,yr after protostar formation, the disk wind is not expected to disappear in a further evolutionary stage, as described above. 
Thus, the disk wind significantly affects the evolution of the disk (or protostellar system).

Since the disk evolution is controlled by the transfer rate of angular momentum, the disk wind should affect the disk evolution. 
To discuss the disk evolution after the main accretion phase, the time evolution of the angular momentum in the disk and outflow is shown in this subsection. 

\begin{figure*}
\begin{center}
\includegraphics[width=0.9\columnwidth]{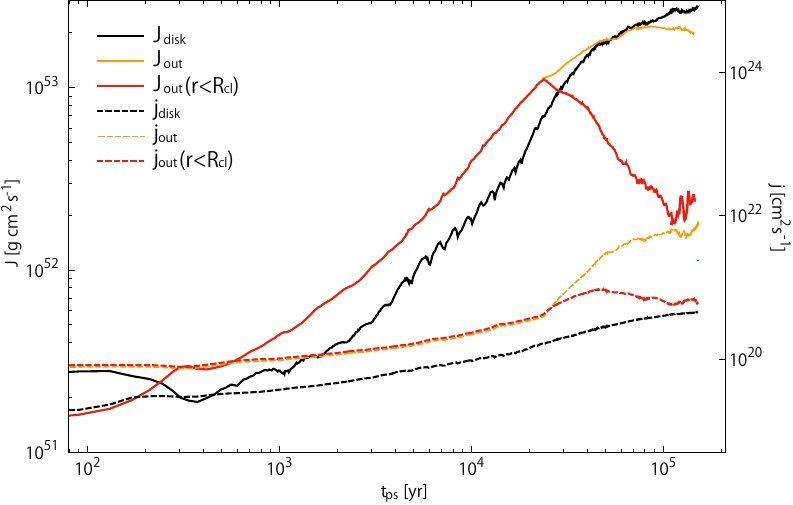}
\end{center}
\caption{
Total (left axis) and specific (right axis) angular momentum of the disk $J_{\rm disk}$, $j_{\rm disk}$ and outflow $J_{\rm out}$, $J_{\rm out} (r<r_{\rm cl})$, $j_{\rm out}$, $j_{\rm out} (r<r_{\rm cl})$ against the elapsed time $t_{\rm ps}$ after protostar formation. 
Orange lines correspond to the (specific) angular momentum of outflow in the whole region $J_{\rm out}$,  $j_{\rm out}$, while  red lines are that within the cloud core $J_{\rm out} (r<r_{\rm cl})$, $j_{\rm out} (r<r_{\rm cl})$. 
}
\label{fig:10}
\end{figure*}

The total and specific angular momenta of the disk and outflow are plotted in Figure~\ref{fig:10}, in which the angular momentum is integrated within the disk and outflow.
Note that the identification of the disk and outflow is described in \S\ref{sec:wholeevo}. 
The total angular momentum of the outflow $J_{\rm out}$ dominates that of the disk $J_{\rm disk}$ until $t_{\rm ps}=2\times10^4$\,yr (solid red line in Fig.~\ref{fig:10}) or $t_{\rm ps} = 5\times 10^4$\,yr (thin orange line in Fig.~\ref{fig:10}). 
A sudden drop of the outflow angular momentum  $J_{\rm out} (r<R_{\rm cl})$ means that the head of the outflow extends outside the star-forming core with a size of $R_{\rm cl}$, as seen in Figure~\ref{fig:1}{\it e}--{\it i}. 
The total angular momentum of the outflow $J_{\rm out}$ is plotted by the orange solid line in Figure~\ref{fig:10}, and includes the gas entrained by the outflow. 
Although there is uncertainty in the total angular momentum of the outflow, its value is comparable to or greater than that of the disk. 
In addition, the specific angular momentum of the outflow within the cloud core ($j_{\rm out} (r<R_{\rm cl})$) is always greater than that of the disk ($j_{\rm disk}$). 
Thus, the outflow has a significant effect on the disk evolution. 

\begin{figure*}
\begin{center}
\includegraphics[width=1.0\columnwidth]{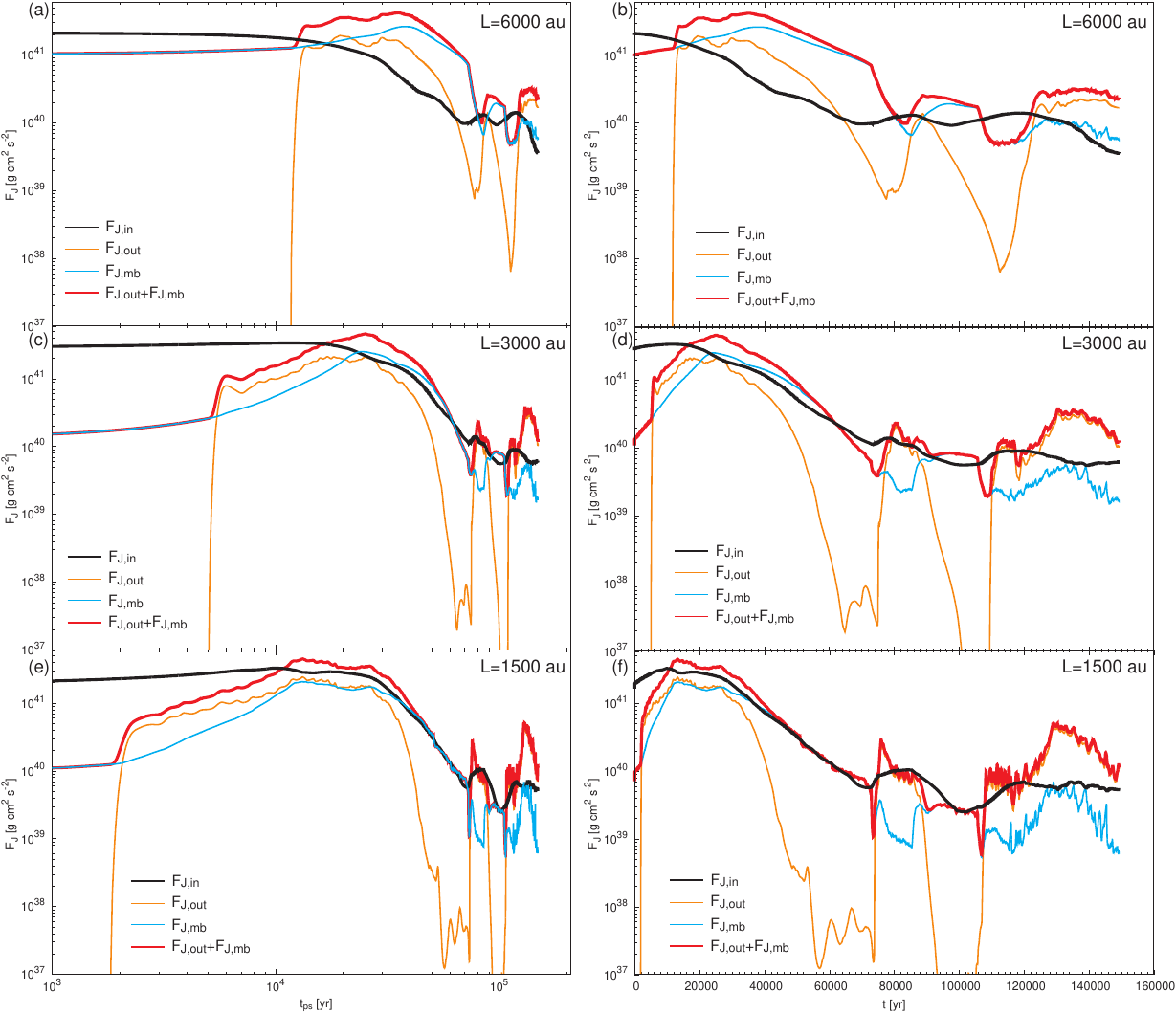}
\end{center}
\caption{
Angular momentum flux inflowing into and outflowing from the box with a size $L=$ 6000\,au (top), 3000\,au (middle) and 1500\,au (bottom) against the elapsed time after protostar formation in logarithmic (left panels) and linear (right panels) scales.
Angular momentum flux of the inflowing gas $F_{\rm J, in}$, outflowing gas $F_{\rm J, out}$ and magnetic braking $F_{\rm J, mb}$ is plotted in each panel. 
}
\label{fig:11}
\end{figure*}

In addition to the outflow, magnetic braking contributes to the angular momentum transfer from the disk. 
To estimate the angular momentum introduced into and transported from the disk, the angular momentum fluxes of the outflow, inflow, and magnetic braking are estimated as 
\begin{equation}
F_{\rm J, out} = \left\vert \int_{v_r > 0} \rho r_c v_\phi  \vect{v}\cdot d\vect{S} \right\vert, 
\label{eq:fout}
\end{equation}
\begin{equation}
F_{\rm J, in} = \left\vert \int_{v_r < 0} \rho r_c v_\phi \vect{v} \cdot d\vect{S} \right\vert,  
\label{eq:fin}
\end{equation}
and
\begin{equation}
F_{\rm J, mb} = \left\vert \int r \frac{B_\phi}{4 \pi} \vect{B} \cdot d\vect{S} \right\vert, 
\label{eq:mb}
\end{equation}
respectively, and are plotted in Figure~\ref{fig:11}.
In equations~(\ref{eq:fout})--(\ref{eq:mb}), the integrals are done on the surface $\vect{S}$ of different grids with grid size of $L=1500$, 3000 and 6000\,au.  
To focus on the early and late phase, the elapsed time $t_{\rm ps}$ is expressed logarithmically (left) and linearly  (right) in Figure~\ref{fig:11}. 
As seen in the left panels of Figure~\ref{fig:11}, an almost constant angular momentum flux $F_{\rm J, in}\sim2\times10^{41}\,\fj$  is introduced into the central region independent of the spatial scale ($L=$ 1500, 3000, 6000\,au). 
Before the outflow reaches each box boundary (1500, 3000, 6000\,au), the angular momentum transport is primarily by the magnetic braking flux $F_{\rm J, mb}$. 
After the outflow reaches each box boundary, the outflow contributes to the angular momentum transfer. 
The angular momentum transferred by the outflow $F_{\rm J, out}$ dominates (or is comparable to) that by the magnetic braking $F_{\rm J, mb}$ after the outflow sufficiently evolves ($t_{\rm ps} \gtrsim 3\times10^4$\,yr). 
However, as described in \S\ref{sec:wholeevo}, the outflow becomes inactive for $3\times10^4\,{\rm yr} \lesssim t_{\rm ps} \lesssim 10^5\,{\rm yr}$, during which the angular momentum transferred by the magnetic braking dominates that by the outflow. 
Note that the outflow is temporarily active at $t_{\rm ps}\sim8\times10^4$\,yr, at which time the angular momentum flux of the outflow $F_{\rm J,out}$ is comparable to that of the magnetic braking $F_{\rm J,mb}$. 

In the left panels of Figure~\ref{fig:11}, for $t_{\rm ps}\gtrsim 5\times 10^4$\,yr, the angular momentum transferred by both outflow and magnetic braking ($F_{\rm J, out}+F_{\rm J,mb}$) is comparable to that introduced into the central region ($F_{\rm J,in}$) on a scale of 1500\,au which covers the whole region of the disk (Fig.~\ref{fig:11}{\it c}). 
In a large scale (Fig.~\ref{fig:11}{\it a}), the angular momentum transferred by the magnetic effects ($F_{\rm J, out}+F_{\rm J,mb}$) is larger than that introduced into the central region ($F_{\rm J,in}$). 

Next, we focus on the evolution of the angular momentum flux in late phase (right panels of Fig.~\ref{fig:11}). 
As described above, the outflow (or disk wind) is activated again as the infalling envelope dissipates because the ram pressure decreases. 
On the other hand, the magnetic braking becomes inefficient as the mass of the infalling envelope decreases.
It is difficult for a less massive envelope to brake a massive disk. 
The low density envelope has a high Alfv\'en speed, which reduces the amplitude of $B_\phi$ at the disk surface that is used in equation (\ref{eq:mb}). 
As shown in Figure~\ref{fig:2}, the disk mass dominates the envelope mass for $t_{\rm ps}\gtrsim5\times10^4$\,yr. 
Thus, after the angular momentum flux $J_{\rm J,mb}$ reaches a peak at $t_{\rm ps} \simeq 2-5\times10^4$\,yr, it gradually decreases (Fig.~\ref{fig:11}). 
Instead of the magnetic braking, the outflow or disk wind can efficiently transfer the angular momentum from the central region.  
For $t_{\rm ps}\gtrsim 10^5$\,yr, the angular momentum flux of the outflow $F_{\rm J,out}$ dominates that of the magnetic braking $F_{\rm J,mb}$. 
In addition, for $t_{\rm ps}\gtrsim 1.2\times10^5$\,yr, the angular momentum flux of the outflow $F_{\rm J,out}$ always dominates those of the magnetic braking $F_{\rm J,mb}$ and the infalling gas $F_{\rm J,in}$ (right panels of Fig.~\ref{fig:11}). 

As described above, since the disk is supported by rotation against gravity of the central protostar, the disk evolution is determined by the transport rate of the angular momentum. 
Finally, we estimate the dissipation timescale of the disk in terms of the angular momentum transport due to the disk wind (or outflow).
As seen in Figure~\ref{fig:10}, the total angular momentum of the disk is about $J_{\rm disk}\simeq3\times 10^{53}\,\Jang$ at the end of the simulation. 
Since the infalling envelope is already mostly dissipated, the disk angular momentum would not significantly increase in the further evolutionary stage. 
On the other hand, the ram pressure cannot suppress the disk wind because the density of the infalling envelope is considerably low. 
As a result, the wind appears in the whole region of the disk, as seen in Figure~\ref{fig:9}. 
Figure~\ref{fig:11} indicates that the angular momentum flux due to the disk wind is $(dJ_{\rm out}/dt=)F_{\rm J,out}\simeq10^{40}\,\fj$ for $t_{\rm ps}\gtrsim 10^5$\,yr.
Assuming constant $F_{\rm J,out}$ in the further evolutionary stage, the timescale to remove the disk angular momentum by the disk wind can be written as
\begin{equation}
t_{\rm diss, J}=\frac{J_{\rm out}}{F_{\rm J,out}} =\frac{J_{\rm out}}{dJ_{\rm out}/dt} \simeq \frac{3\times10^{53}}{10^{40}} \simeq 3\times10^
{13}\,{\rm s} = 9.5\times10^5\,\rm{yr}. 
\end{equation}
Thus, although there are many uncertainties, we can estimate the disk dissipation timescale as $t_{\rm diss,J}\sim10^6$\,yr. 

\subsection{Mass Dispersal by Disk Wind}
\begin{figure*}
\begin{center}
\includegraphics[width=0.9\columnwidth]{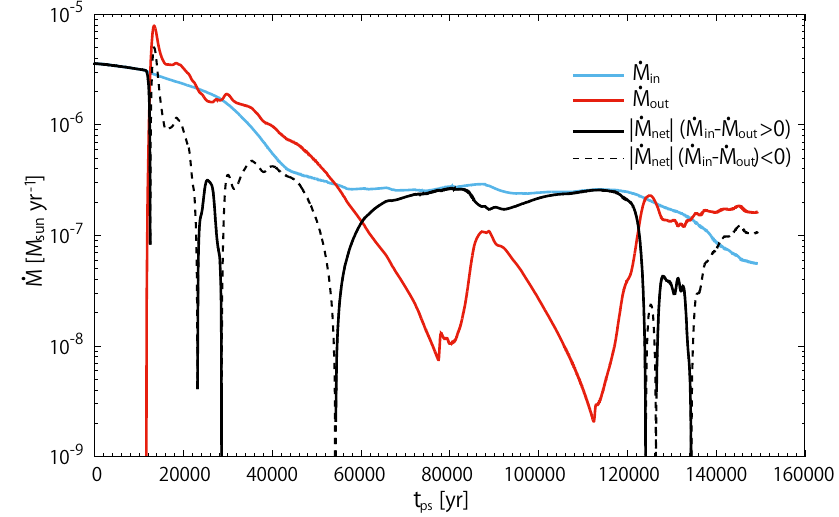}
\end{center}
\caption{
Mass inflow $\dot{M}_{\rm in}$ and outflow $\dot{M}_{\rm out}$ rate on the surface of the box with a size of $3000$\,au versus $t_{\rm ps}$, the elapsed time after protostar formation. 
The net inflow or outflow rate $\vert \dot{M}_{\rm net} \vert$ are also plotted.
}
\label{fig:12}
\end{figure*}

We discuss the dispersal timescale of the disk mass due to the disk wind assuming a constant mass loss rate. 
To investigate the net mass loss rate, the inflow $\dot{M}_{\rm in}$ and outflow $\dot{M}_{\rm out}$ mass rates on the surface of the $l=6$ grid with a side length of $3000$\,au are plotted in Figure~\ref{fig:12}. 
These rates are estimated as 
\begin{equation}
\dot{M}_{\rm in} (l=6) = \int_{\rm surface \, of \, {\rm l=6} \ grid} \rho \, \vect{v} \cdot \vect{n} (<0) \, dS,
\end{equation}
where $\vect{n}$ is the normal vector of each surface and we executed the integration only when $\vect{v} \cdot \vect{n}<0$ (i.e., inflow).
The mass outflow rate is calculated as  
\begin{equation}
\dot{M}_{\rm out} (l=6) = \int_{\rm surface \, of \, level\, {\rm l=6} \ grid} \rho \, \vect{v} \cdot \vect{n} (>0) \, dS.
\end{equation}
As seen in Figure~\ref{fig:1}, since the diameter of the initial cloud is $1.2 \times 10^4$\,au, we estimated the inflow and outflow mass rate at the surface of the grid with the size of the half initial cloud diameter. 

Figure~\ref{fig:12} shows that the mass inflow rate $\dot{M}_{\rm in}$ gradually decreases from $3\times10^{-6}$ to $6\times10^{-8}\,\msunyr$ by the end of the simulation. 
On the other hand, after the mass outflow rate continues to decrease until $t_{\rm ps}\sim 8\times10^4$\,yr during which the outflow mass loss rate 
 sometimes exceeds the mass infall rate for $t_{\rm ps}\lesssim 6\times10^4$\,yr. 
Then, although once the outflow mass loss rate increases for $t_{\rm ps} \simeq 8\times10^4-1.1\times10^5$\,yr, the mass loss rate never exceeds the mass infall rate in the range of $ 6\times 10^4\,{\rm yr} \lesssim t_{\rm ps} \lesssim 1.2\times 10^5\,{\rm yr}$. 
On the other hand, for $t_{\rm ps}\gtrsim 1.3\times10^5$\,yr, the mass loss rate exceeds the mass inflow rate until the end of the simulation. 

As described in \S\ref{sec:outflowdriving},  the outflow or disk wind continues to be driven by the circumstellar disk after the infalling envelope dissipates. 
As shown in Figure~\ref{fig:2}, the disk mass is $M_{\rm disk}\simeq 0.3\,\msun$ at the end of the simulation. 
The net mass loss rate, which is described by the dotted line in Figure~\ref{fig:12}, is $\dot{M}_{\rm out}\simeq 10^{-7}\,\msunyr$ at the end of the simulation. 
Thus, the disk dispersal timescale is estimated as $t_{\rm diss} \sim M_{\rm disk} / \dot{M}_{\rm net} \sim 3\times10^6$\,yr. 
Therefore, in terms of the mass, the disk would disappear due to the disk wind within $\sim10^6$\,yr. 

\subsection{Effects of External Medium and Outer Boundary Condition}
\label{sec:numericaleffects}
Finally, we discuss the effects of the boundary condition and the external medium distributed outside the cloud. 
We need to carefully consider the boundary condition of the magnetic field in MHD simulations. 
We adopted a fixed boundary condition of the magnetic field, as described in \S\ref{sec:settings}. 
The boundary condition of the magnetic field is imposed on the surface of the $l=1$ grid which is 
$0.98\times10^5$\,au ($L(1)/2=(1.96\times10^5)/2\,{\rm au})$ far from the center of the collapsing cloud. 
The Alfv\'en velocity $v_{\rm A,ext}$ in the region outside the cloud is $v_{\rm A,ext} = B_0/\sqrt{4\pi\, \rho_{\rm ISM}} = 0.25\,\kms$. 
Thus, it takes $1.9\times10^6$\,yr $(=0.98\times10^5\,{\rm au}/0.25\,\kms)$ for the Alfv\'en wave from the cloud center to reach the computational boundary on the surface of the $l=1$ grid.
Thus, the Alfv\'en wave generated from the cloud center never reaches the computational boundary within the calculation time $\sim2\times10^5$\,yr. 
In order to minimize the influence of the boundary as much as possible, we imposed this boundary condition and large domain outside the cloud.

As described in \S\ref{sec:settings}, a gas with density $\rho_{\rm ext}=3.4\times 10^{-19}$\,{\rm g}\,{\rm cm}$^{-3}$ ($n_{\rm ext}=8.5\times10^4\,\cc$) is uniformly distributed outside the star-forming cloud core $r>R_{\rm cl}$. 
As shown in Figure~\ref{fig:1}, the gas density decreases within the cloud ($r<R_{\rm cl}$) and is lower than that outside the core ($r>R_{\rm cl}$) in the late accretion stage ($t_{\rm ps} \gtrsim 5\times10^4$\,yr). 
Since the circumstellar disk is connected to the medium outside the star-forming core through magnetic fields, the magnetic braking works between the external medium and circumstellar disk. 
To evaluate the efficiency of the magnetic braking, we estimate the mass of the cylinder above the circumstellar disk \citep{mouschovias79,mouschovias80,hirano20}.
Considering the cylinder above the disk with a radius of 300\,au and assuming a disk formation timescale of $10^5$\,yr, the mass of the cylinder corresponding to matter affected by Alfv\'en waves is estimated as 
 \begin{equation}
 M_{\rm cyl} = \pi (300\, {\rm au})^2 \times (v_{\rm A,ext} \times 10^5\,{\rm yr})\times \rho_{\rm ext} = 8.2\times10^{-4}\,\msun. 
 \end{equation}
Thus, $M_{\rm cyl}$ is much less than the disk mass ($\sim 0.01-0.1\,\msun$, see Fig.~\ref{fig:2}), indicating that it is difficult for the external medium to brake the circumstellar disk. 
Therefore, we expect that magnetic braking from the external medium does not significantly affect the disk.

\section{Summary}
\label{sec:summary}
We performed a very long-term simulation of star and disk formation using a three-dimensional nonideal MHD nested grid code. 
As the initial state, we prepared the prestellar cloud having a critical Bonnor-Ebert density profile.
After starting the calculation, we created the sink cell when the central number density exceeded $n_{\rm thr}=10^{13}\,\cc$. 
We calculated the evolution of the cloud until about $1.5\times10^5$\,yr after protostar formation (or sink creation). 
At the end of the simulation, the mass ratio of envelope ($M_{\rm env}$), protostar ($M_{\rm ps}$), circumstellar disk ($M_{\rm env}$), and outflow ($M_{\rm out}$) to the initial cloud mass ($M_{\rm cl}$) are $M_{\rm env}/M_{\rm cl}=0.02$, $M_{\rm ps}/M_{\rm cl}=0.44$, $M_{\rm disk}/M_{\rm cl}=0.24$ and $M_{\rm out}/M_{\rm cl}=0.30$, respectively. 
Thus, the infalling envelope dissipates around the disk and protostar at the end of the simulation.

After the first core formation prior to the protostar formation, the magnetically driven outflow appears. 
The outflow creates a cavity-like structure in the star-forming cloud core. 
Although the outflow is very strong just after protostar formation, it gradually weakens in the main accretion phase.
Since the outflow is powered by the infall which has a decreasing mass accretion rate, it decreases in strength and eventually disappears. 
However, a large fraction of mass is ejected by the outflow before its (temporary) disappearance. 
Since the mass of the star-forming cloud core decreases and gravitational potential of the cloud becomes shallow, the outer part of the cloud slowly moves outward due to the pressure gradient and magnetic forces. 
This happens especially on the equatorial plane, where the pressure gradient force is strong.
A part of the infalling gas is ejected by the magnetically and thermally driven outflow, while the remainder falls onto the circumstellar disk and protostar. 
As a result, the envelope dissipates and the gas density around the disk considerably decreases. 

The density of the infalling envelope becomes low as the envelope dissipates, resulting in a weakening of the ram pressure above and below the disk. 
Then, the outflow (as a disk wind) is activated again. 
The wind appears in the whole disk region in the late accretion phase during which the mass of the envelope is as low as $<0.1\,\msun$.  

The disk radius is $r_{\rm disk}\lesssim 50$\,au in the main accretion phase, during which magnetic braking suppresses the disk growth. 
In addition to the magnetic braking, the outflow also transports the angular momentum from the star-forming core to the interstellar medium. 
The magnetic braking is ineffective as the envelope dissipates in the late accretion phase, because a less massive envelope cannot brake a more massive rotating disk. 
In the intermediate accretion stage where the mass of the infalling envelope is comparable to the disk mass or the protostellar mass, the disk gravitational instability develops to form a spiral structure. 
However, as the infalling envelope dissipates, the mass accretion rate onto the disk decreases. 
Then, the Toomre $Q$ parameter also gradually increases and a stable disk with size of $\gtrsim 300$\,au remains at the end of the simulation. 

Finally, we estimated the dissipation or dispersal timescale of the circumstellar disk. 
The disk wind should be driven by the disk after the infalling envelope dissipates. 
Thus, the disk angular momentum is transported by the disk wind. 
Using the disk angular momentum $J_{\rm disk}$ and angular momentum flux rate $dJ_{\rm out}/dt$ at the end of the simulation, we estimated the dissipation timescale of the disk as $t_{\rm diss}=J_{\rm disk}/(dJ_{\rm out}/dt)\simeq 10^6$\,yr. 
We also estimated the mass dispersal timescale of the disk by the disk wind using the disk mass $M_{\rm disk}$ and the mass loss rate of the disk wind $\dot{M}_{\rm out}$. 
The disk dispersal timescale is $t_{\rm disp}\sim M_{\rm disk}/\dot{M}_{\rm out}\sim 3\times 10^6$\,yr. 
Although more time integration is necessary to make a final conclusion, our global simulation implies that the disk lifetime is about $10^6$\,yr. 

\section*{Acknowledgements}
This research used the computational resources of the HPCI system provided by the Cyber Science Center at Tohoku University and the Cybermedia Center at Osaka University (Project ID: hp200004, hp210004, hp220003, hp230035, hp240010).
Simulations reported in this paper were also performed by 2020, 2021, 2022, 2023, 2024 Koubo Kadai on Earth Simulator (NEC SX-ACE) at JAMSTEC. 
The present study was supported by JSPS KAKENHI Grant (JP21H00046, JP21K03617: MNM).
This work was supported by a NAOJ ALMA Scientific Research grant (No. 2022-22B). 
S.B. was supported by a Discovery Grant from NSERC.

\bibliography{machidabib}{}

\begin{thebibliography}{}
\expandafter\ifx\csname natexlab\endcsname\relax\def\natexlab#1{#1}\fi
\providecommand{\url}[1]{\href{#1}{#1}}
\providecommand{\dodoi}[1]{doi:~\href{http://doi.org/#1}{\nolinkurl{#1}}}
\providecommand{\doeprint}[1]{\href{http://ascl.net/#1}{\nolinkurl{http://ascl.net/#1}}}
\providecommand{\doarXiv}[1]{\href{https://arxiv.org/abs/#1}{\nolinkurl{https://arxiv.org/abs/#1}}}

\bibitem[{{Alves} {et~al.}(2017){Alves}, {Girart}, {Caselli}, {Franco}, {Zhao},
  {Vlemmings}, {Evans}, \& {Ricci}}]{alves17}
{Alves}, F.~O., {Girart}, J.~M., {Caselli}, P., {et~al.} 2017, \aap, 603, L3,
  \dodoi{10.1051/0004-6361/201731077}

\bibitem[{{Andrews} {et~al.}(2018){Andrews}, {Huang}, {P{\'e}rez}, {Isella},
  {Dullemond}, {Kurtovic}, {Guzm{\'a}n}, {Carpenter}, {Wilner}, {Zhang}, {Zhu},
  {Birnstiel}, {Bai}, {Benisty}, {Hughes}, {{\"O}berg}, \& {Ricci}}]{andrews18}
{Andrews}, S.~M., {Huang}, J., {P{\'e}rez}, L.~M., {et~al.} 2018, \apjl, 869,
  L41, \dodoi{10.3847/2041-8213/aaf741}

\bibitem[{{Aso} \& {Machida}(2020)}]{aso20}
{Aso}, Y., \& {Machida}, M.~N. 2020, \apj, 905, 174,
  \dodoi{10.3847/1538-4357/abc6fc}

\bibitem[{{Aso} {et~al.}(2015){Aso}, {Ohashi}, {Saigo}, {Koyamatsu}, {Aikawa},
  {Hayashi}, {Machida}, {Saito}, {Takakuwa}, {Tomida}, {Tomisaka}, \&
  {Yen}}]{aso15}
{Aso}, Y., {Ohashi}, N., {Saigo}, K., {et~al.} 2015, \apj, 812, 27,
  \dodoi{10.1088/0004-637X/812/1/27}

\bibitem[{{Aso} {et~al.}(2019){Aso}, {Hirano}, {Aikawa}, {Machida}, {Ohashi},
  {Saito}, {Takakuwa}, {Yen}, \& {Williams}}]{aso19}
{Aso}, Y., {Hirano}, N., {Aikawa}, Y., {et~al.} 2019, \apj, 887, 209,
  \dodoi{10.3847/1538-4357/ab5284}

\bibitem[{{Bai}(2015)}]{bai15}
{Bai}, X.-N. 2015, \apj, 798, 84, \dodoi{10.1088/0004-637X/798/2/84}

\bibitem[{{Bai} \& {Stone}(2013)}]{bai13}
{Bai}, X.-N., \& {Stone}, J.~M. 2013, \apj, 769, 76,
  \dodoi{10.1088/0004-637X/769/1/76}

\bibitem[{{Balbus}(2003)}]{balbus03}
{Balbus}, S.~A. 2003, \araa, 41, 555,
  \dodoi{10.1146/annurev.astro.41.081401.155207}

\bibitem[{{Balbus} \& {Hawley}(1998)}]{Balbus98}
{Balbus}, S.~A., \& {Hawley}, J.~F. 1998, Reviews of Modern Physics, 70, 1,
  \dodoi{10.1103/RevModPhys.70.1}

\bibitem[{{Banerjee} \& {Pudritz}(2006)}]{banerjee06}
{Banerjee}, R., \& {Pudritz}, R.~E. 2006, \apj, 641, 949,
  \dodoi{10.1086/500496}

\bibitem[{{Basu} {et~al.}(2024){Basu}, {Sharkawi}, \& {Machida}}]{Basu24}
{Basu}, S., {Sharkawi}, M., \& {Machida}, M.~N. 2024, \apj, 964, 116,
  \dodoi{10.3847/1538-4357/ad1bf3}

\bibitem[{{Bjerkeli} {et~al.}(2016){Bjerkeli}, {van der Wiel}, {Harsono},
  {Ramsey}, \& {J{\o}rgensen}}]{bjerkeli16}
{Bjerkeli}, P., {van der Wiel}, M. H.~D., {Harsono}, D., {Ramsey}, J.~P., \&
  {J{\o}rgensen}, J.~K. 2016, \nat, 540, 406, \dodoi{10.1038/nature20600}

\bibitem[{{Blandford} \& {Payne}(1982)}]{blandford82}
{Blandford}, R.~D., \& {Payne}, D.~G. 1982, \mnras, 199, 883,
  \dodoi{10.1093/mnras/199.4.883}

\bibitem[{{Booth} {et~al.}(2021){Booth}, {Tabone}, {Ilee}, {Walsh}, {Aikawa},
  {Andrews}, {Bae}, {Bergin}, {Bergner}, {Bosman}, {Calahan}, {Cataldi},
  {Cleeves}, {Czekala}, {Guzm{\'a}n}, {Huang}, {Law}, {Le Gal}, {Long},
  {Loomis}, {M{\'e}nard}, {Nomura}, {{\"O}berg}, {Qi}, {Schwarz}, {Teague},
  {Tsukagoshi}, {Wilner}, {Yamato}, \& {Zhang}}]{booth21}
{Booth}, A.~S., {Tabone}, B., {Ilee}, J.~D., {et~al.} 2021, \apjs, 257, 16,
  \dodoi{10.3847/1538-4365/ac1ad4}

\bibitem[{{Cieza} {et~al.}(2019){Cieza}, {Ru{\'\i}z-Rodr{\'\i}guez}, {Hales},
  {Casassus}, {P{\'e}rez}, {Gonzalez-Ruilova}, {C{\'a}novas}, {Williams},
  {Zurlo}, {Ansdell}, {Avenhaus}, {Bayo}, {Bertrang}, {Christiaens}, {Dent},
  {Ferrero}, {Gamen}, {Olofsson}, {Orcajo}, {Pe{\~n}a Ram{\'\i}rez},
  {Principe}, {Schreiber}, \& {van der Plas}}]{cieza19}
{Cieza}, L.~A., {Ru{\'\i}z-Rodr{\'\i}guez}, D., {Hales}, A., {et~al.} 2019,
  \mnras, 482, 698, \dodoi{10.1093/mnras/sty2653}

\bibitem[{{Dapp} {et~al.}(2012){Dapp}, {Basu}, \& {Kunz}}]{dapp12}
{Dapp}, W.~B., {Basu}, S., \& {Kunz}, M.~W. 2012, \aap, 541, A35,
  \dodoi{10.1051/0004-6361/201117876}

\bibitem[{{de Valon} {et~al.}(2020){de Valon}, {Dougados}, {Cabrit}, {Louvet},
  {Zapata}, \& {Mardones}}]{devalon20}
{de Valon}, A., {Dougados}, C., {Cabrit}, S., {et~al.} 2020, \aap, 634, L12,
  \dodoi{10.1051/0004-6361/201936950}

\bibitem[{{Dullemond} {et~al.}(2018){Dullemond}, {Birnstiel}, {Huang},
  {Kurtovic}, {Andrews}, {Guzm{\'a}n}, {P{\'e}rez}, {Isella}, {Zhu}, {Benisty},
  {Wilner}, {Bai}, {Carpenter}, {Zhang}, \& {Ricci}}]{dullemond18}
{Dullemond}, C.~P., {Birnstiel}, T., {Huang}, J., {et~al.} 2018, \apjl, 869,
  L46, \dodoi{10.3847/2041-8213/aaf742}

\bibitem[{{Fang} {et~al.}(2023){Fang}, {Wang}, {Herczeg}, {Hashimoto}, {Xu},
  {Nemer}, {Pascucci}, {Haffert}, \& {Aoyama}}]{fang23}
{Fang}, M., {Wang}, L., {Herczeg}, G.~J., {et~al.} 2023, Nature Astronomy, 7,
  905, \dodoi{10.1038/s41550-023-02004-x}

\bibitem[{{Fern{\'a}ndez-L{\'o}pez} {et~al.}(2020){Fern{\'a}ndez-L{\'o}pez},
  {Zapata}, {Rodr{\'\i}guez}, {Vazzano}, {Guzm{\'a}n}, \& {L{\'o}pez}}]{Fern20}
{Fern{\'a}ndez-L{\'o}pez}, M., {Zapata}, L.~A., {Rodr{\'\i}guez}, L.~F.,
  {et~al.} 2020, \aj, 159, 171, \dodoi{10.3847/1538-3881/ab7a10}

\bibitem[{{Flaherty} {et~al.}(2020){Flaherty}, {Hughes}, {Simon}, {Qi}, {Bai},
  {Bulatek}, {Andrews}, {Wilner}, \& {K{\'o}sp{\'a}l}}]{flaherty20}
{Flaherty}, K., {Hughes}, A.~M., {Simon}, J.~B., {et~al.} 2020, \apj, 895, 109,
  \dodoi{10.3847/1538-4357/ab8cc5}

\bibitem[{{Flaherty} {et~al.}(2015){Flaherty}, {Hughes}, {Rosenfeld},
  {Andrews}, {Chiang}, {Simon}, {Kerzner}, \& {Wilner}}]{flaherty15}
{Flaherty}, K.~M., {Hughes}, A.~M., {Rosenfeld}, K.~A., {et~al.} 2015, \apj,
  813, 99, \dodoi{10.1088/0004-637X/813/2/99}

\bibitem[{{Gressel} {et~al.}(2015){Gressel}, {Turner}, {Nelson}, \&
  {McNally}}]{gressel15}
{Gressel}, O., {Turner}, N.~J., {Nelson}, R.~P., \& {McNally}, C.~P. 2015,
  \apj, 801, 84, \dodoi{10.1088/0004-637X/801/2/84}

\bibitem[{{Hara} {et~al.}(2013){Hara}, {Shimajiri}, {Tsukagoshi}, {Kurono},
  {Saigo}, {Nakamura}, {Saito}, {Wilner}, \& {Kawabe}}]{hara13}
{Hara}, C., {Shimajiri}, Y., {Tsukagoshi}, T., {et~al.} 2013, \apj, 771, 128,
  \dodoi{10.1088/0004-637X/771/2/128}

\bibitem[{{Harada} {et~al.}(2023){Harada}, {Tokuda}, {Yamasaki}, {Sato},
  {Omura}, {Hirano}, {Onishi}, {Tachihara}, \& {Machida}}]{harada23}
{Harada}, N., {Tokuda}, K., {Yamasaki}, H., {et~al.} 2023, \apj, 945, 63,
  \dodoi{10.3847/1538-4357/acb930}

\bibitem[{{Hennebelle} \& {Fromang}(2008)}]{hennebelle08}
{Hennebelle}, P., \& {Fromang}, S. 2008, \aap, 477, 9,
  \dodoi{10.1051/0004-6361:20078309}

\bibitem[{{Hirano} {et~al.}(2020){Hirano}, {Tsukamoto}, {Basu}, \&
  {Machida}}]{hirano20}
{Hirano}, S., {Tsukamoto}, Y., {Basu}, S., \& {Machida}, M.~N. 2020, \apj, 898,
  118, \dodoi{10.3847/1538-4357/ab9f9d}

\bibitem[{{Hirota} {et~al.}(2017){Hirota}, {Machida}, {Matsushita}, {Motogi},
  {Matsumoto}, {Kim}, {Burns}, \& {Honma}}]{hirota17}
{Hirota}, T., {Machida}, M.~N., {Matsushita}, Y., {et~al.} 2017, Nature
  Astronomy, 1, 0146, \dodoi{10.1038/s41550-017-0146}

\bibitem[{{Huang} {et~al.}(2021){Huang}, {Bergin}, {{\"O}berg}, {Andrews},
  {Teague}, {Law}, {Kalas}, {Aikawa}, {Bae}, {Bergner}, {Booth}, {Bosman},
  {Calahan}, {Cataldi}, {Cleeves}, {Czekala}, {Ilee}, {Le Gal}, {Guzm{\'a}n},
  {Long}, {Loomis}, {M{\'e}nard}, {Nomura}, {Qi}, {Schwarz}, {Tsukagoshi},
  {van't Hoff}, {Walsh}, {Wilner}, {Yamato}, \& {Zhang}}]{huang21}
{Huang}, J., {Bergin}, E.~A., {{\"O}berg}, K.~I., {et~al.} 2021, \apjs, 257,
  19, \dodoi{10.3847/1538-4365/ac143e}

\bibitem[{{Joos} {et~al.}(2012){Joos}, {Hennebelle}, \& {Ciardi}}]{joos12}
{Joos}, M., {Hennebelle}, P., \& {Ciardi}, A. 2012, \aap, 543, A128,
  \dodoi{10.1051/0004-6361/201118730}

\bibitem[{{Kawasaki} {et~al.}(2021){Kawasaki}, {Koga}, \&
  {Machida}}]{kawasaki21}
{Kawasaki}, Y., {Koga}, S., \& {Machida}, M.~N. 2021, \mnras, 504, 5588,
  \dodoi{10.1093/mnras/stab1224}

\bibitem[{{Koga} {et~al.}(2022){Koga}, {Kawasaki}, \& {Machida}}]{koga22}
{Koga}, S., {Kawasaki}, Y., \& {Machida}, M.~N. 2022, \mnras, 515, 6073,
  \dodoi{10.1093/mnras/stac2115}

\bibitem[{{Koga} \& {Machida}(2023)}]{koga23}
{Koga}, S., \& {Machida}, M.~N. 2023, \mnras, 519, 3595,
  \dodoi{10.1093/mnras/stac3503}

\bibitem[{{Larson}(1969)}]{larson69}
{Larson}, R.~B. 1969, \mnras, 145, 271, \dodoi{10.1093/mnras/145.3.271}

\bibitem[{{Launhardt} {et~al.}(2023){Launhardt}, {Pavlyuchenkov}, {Akimkin},
  {Dutrey}, {Gueth}, {Guilloteau}, {Henning}, {Pi{\'e}tu}, {Schreyer},
  {Semenov}, {Stecklum}, \& {Bourke}}]{launhardt23}
{Launhardt}, R., {Pavlyuchenkov}, Y.~N., {Akimkin}, V.~V., {et~al.} 2023, \aap,
  678, A135, \dodoi{10.1051/0004-6361/202347483}

\bibitem[{{Lee} {et~al.}(2014){Lee}, {Hirano}, {Zhang}, {Shang}, {Ho}, \&
  {Krasnopolsky}}]{Lee14}
{Lee}, C.-F., {Hirano}, N., {Zhang}, Q., {et~al.} 2014, \apj, 786, 114,
  \dodoi{10.1088/0004-637X/786/2/114}

\bibitem[{{Lee} {et~al.}(2017){Lee}, {Ho}, {Li}, {Hirano}, {Zhang}, \&
  {Shang}}]{Lee17}
{Lee}, C.-F., {Ho}, P. T.~P., {Li}, Z.-Y., {et~al.} 2017, Nature Astronomy, 1,
  0152, \dodoi{10.1038/s41550-017-0152}

\bibitem[{{L{\'o}pez-V{\'a}zquez} {et~al.}(2023){L{\'o}pez-V{\'a}zquez},
  {Zapata}, \& {Lee}}]{lopez23}
{L{\'o}pez-V{\'a}zquez}, J.~A., {Zapata}, L.~A., \& {Lee}, C.-F. 2023, \apj,
  944, 63, \dodoi{10.3847/1538-4357/acb439}

\bibitem[{{Louvet} {et~al.}(2018){Louvet}, {Dougados}, {Cabrit}, {Mardones},
  {M{\'e}nard}, {Tabone}, {Pinte}, \& {Dent}}]{louvet18}
{Louvet}, F., {Dougados}, C., {Cabrit}, S., {et~al.} 2018, \aap, 618, A120,
  \dodoi{10.1051/0004-6361/201731733}

\bibitem[{{Machida}(2021)}]{machida21}
{Machida}, M.~N. 2021, \mnras, 508, 3208, \dodoi{10.1093/mnras/stab2626}

\bibitem[{{Machida} \& {Basu}(2019)}]{machida19}
{Machida}, M.~N., \& {Basu}, S. 2019, \apj, 876, 149,
  \dodoi{10.3847/1538-4357/ab18a7}

\bibitem[{{Machida} \& {Hosokawa}(2013)}]{machida13}
{Machida}, M.~N., \& {Hosokawa}, T. 2013, \mnras, 431, 1719,
  \dodoi{10.1093/mnras/stt291}

\bibitem[{{Machida} \& {Hosokawa}(2020)}]{machida20}
---. 2020, \mnras, 499, 4490, \dodoi{10.1093/mnras/staa3139}

\bibitem[{{Machida} {et~al.}(2006){Machida}, {Inutsuka}, \&
  {Matsumoto}}]{machida06}
{Machida}, M.~N., {Inutsuka}, S.-i., \& {Matsumoto}, T. 2006, \apjl, 647, L151,
  \dodoi{10.1086/507179}

\bibitem[{{Machida} {et~al.}(2007){Machida}, {Inutsuka}, \&
  {Matsumoto}}]{machida07}
---. 2007, \apj, 670, 1198, \dodoi{10.1086/521779}

\bibitem[{{Machida} {et~al.}(2009){Machida}, {Inutsuka}, \&
  {Matsumoto}}]{machida09}
---. 2009, \apjl, 699, L157, \dodoi{10.1088/0004-637X/699/2/L157}

\bibitem[{{Machida} {et~al.}(2010){Machida}, {Inutsuka}, \&
  {Matsumoto}}]{machida10}
---. 2010, \apj, 724, 1006, \dodoi{10.1088/0004-637X/724/2/1006}

\bibitem[{{Machida} {et~al.}(2014){Machida}, {Inutsuka}, \&
  {Matsumoto}}]{machida14}
---. 2014, \mnras, 438, 2278, \dodoi{10.1093/mnras/stt2343}

\bibitem[{{Machida} \& {Matsumoto}(2011)}]{machida11}
{Machida}, M.~N., \& {Matsumoto}, T. 2011, \mnras, 413, 2767,
  \dodoi{10.1111/j.1365-2966.2011.18349.x}

\bibitem[{{Machida} \& {Matsumoto}(2012)}]{machida12}
---. 2012, \mnras, 421, 588, \dodoi{10.1111/j.1365-2966.2011.20336.x}

\bibitem[{{Machida} {et~al.}(2005){Machida}, {Matsumoto}, {Tomisaka}, \&
  {Hanawa}}]{machida05a}
{Machida}, M.~N., {Matsumoto}, T., {Tomisaka}, K., \& {Hanawa}, T. 2005,
  \mnras, 362, 369, \dodoi{10.1111/j.1365-2966.2005.09297.x}

\bibitem[{{Machida} {et~al.}(2004){Machida}, {Tomisaka}, \&
  {Matsumoto}}]{machida04}
{Machida}, M.~N., {Tomisaka}, K., \& {Matsumoto}, T. 2004, \mnras, 348, L1,
  \dodoi{10.1111/j.1365-2966.2004.07402.x}

\bibitem[{{Maret} {et~al.}(2020){Maret}, {Maury}, {Belloche}, {Gaudel},
  {Andr{\'e}}, {Cabrit}, {Codella}, {Lef{\'e}vre}, {Podio}, {Anderl}, {Gueth},
  \& {Hennebelle}}]{maret20}
{Maret}, S., {Maury}, A.~J., {Belloche}, A., {et~al.} 2020, \aap, 635, A15,
  \dodoi{10.1051/0004-6361/201936798}

\bibitem[{{Masunaga} \& {Inutsuka}(2000)}]{masunaga00}
{Masunaga}, H., \& {Inutsuka}, S.-i. 2000, \apj, 531, 350,
  \dodoi{10.1086/308439}

\bibitem[{{Matsushita} {et~al.}(2017){Matsushita}, {Machida}, {Sakurai}, \&
  {Hosokawa}}]{matsushita17}
{Matsushita}, Y., {Machida}, M.~N., {Sakurai}, Y., \& {Hosokawa}, T. 2017,
  \mnras, 470, 1026, \dodoi{10.1093/mnras/stx893}

\bibitem[{{Matsushita} {et~al.}(2021){Matsushita}, {Takahashi}, {Ishii},
  {Tomisaka}, {Ho}, {Carpenter}, \& {Machida}}]{matsushita21}
{Matsushita}, Y., {Takahashi}, S., {Ishii}, S., {et~al.} 2021, \apj, 916, 23,
  \dodoi{10.3847/1538-4357/ac069f}

\bibitem[{{McKee} \& {Ostriker}(2007)}]{mcKee07}
{McKee}, C.~F., \& {Ostriker}, E.~C. 2007, \araa, 45, 565,
  \dodoi{10.1146/annurev.astro.45.051806.110602}

\bibitem[{{Mouschovias} \& {Paleologou}(1979)}]{mouschovias79}
{Mouschovias}, T.~C., \& {Paleologou}, E.~V. 1979, \apj, 230, 204,
  \dodoi{10.1086/157077}

\bibitem[{{Mouschovias} \& {Paleologou}(1980)}]{mouschovias80}
---. 1980, \apj, 237, 877, \dodoi{10.1086/157936}

\bibitem[{{Murillo} {et~al.}(2013){Murillo}, {Lai}, {Bruderer}, {Harsono}, \&
  {van Dishoeck}}]{murillo13}
{Murillo}, N.~M., {Lai}, S.-P., {Bruderer}, S., {Harsono}, D., \& {van
  Dishoeck}, E.~F. 2013, \aap, 560, A103, \dodoi{10.1051/0004-6361/201322537}

\bibitem[{{Nakano} {et~al.}(1995){Nakano}, {Hasegawa}, \& {Norman}}]{nakano95}
{Nakano}, T., {Hasegawa}, T., \& {Norman}, C. 1995, \apj, 450, 183,
  \dodoi{10.1086/176130}

\bibitem[{{{\"O}berg} {et~al.}(2021){{\"O}berg}, {Guzm{\'a}n}, {Walsh},
  {Aikawa}, {Bergin}, {Law}, {Loomis}, {Alarc{\'o}n}, {Andrews}, {Bae},
  {Bergner}, {Boehler}, {Booth}, {Bosman}, {Calahan}, {Cataldi}, {Cleeves},
  {Czekala}, {Furuya}, {Huang}, {Ilee}, {Kurtovic}, {Le Gal}, {Liu}, {Long},
  {M{\'e}nard}, {Nomura}, {P{\'e}rez}, {Qi}, {Schwarz}, {Sierra}, {Teague},
  {Tsukagoshi}, {Yamato}, {van't Hoff}, {Waggoner}, {Wilner}, \&
  {Zhang}}]{oberg21}
{{\"O}berg}, K.~I., {Guzm{\'a}n}, V.~V., {Walsh}, C., {et~al.} 2021, \apjs,
  257, 1, \dodoi{10.3847/1538-4365/ac1432}

\bibitem[{{Ohashi} {et~al.}(2014){Ohashi}, {Saigo}, {Aso}, {Aikawa},
  {Koyamatsu}, {Machida}, {Saito}, {Takahashi}, {Takakuwa}, {Tomida},
  {Tomisaka}, \& {Yen}}]{ohashi14}
{Ohashi}, N., {Saigo}, K., {Aso}, Y., {et~al.} 2014, \apj, 796, 131,
  \dodoi{10.1088/0004-637X/796/2/131}

\bibitem[{{Ohashi} {et~al.}(2023){Ohashi}, {Tobin}, {J{\o}rgensen}, {Takakuwa},
  {Sheehan}, {Aikawa}, {Li}, {Looney}, {Williams}, {Aso}, {Sharma}, {Sai},
  {Yamato}, {Lee}, {Tomida}, {Yen}, {Encalada}, {Flores}, {Gavino}, {Kido},
  {Han}, {Lin}, {Narayanan}, {Phuong}, {Santamar{\'\i}a-Miranda}, {Thieme},
  {van't Hoff}, {de Gregorio-Monsalvo}, {Koch}, {Kwon}, {Lai}, {Lee},
  {Plunkett}, {Saigo}, {Hirano}, {Lam}, \& {Mori}}]{ohashi23}
{Ohashi}, N., {Tobin}, J.~J., {J{\o}rgensen}, J.~K., {et~al.} 2023, \apj, 951,
  8, \dodoi{10.3847/1538-4357/acd384}

\bibitem[{{Omura} {et~al.}(2024){Omura}, {Tokuda}, \& {Machida}}]{omura24}
{Omura}, M., {Tokuda}, K., \& {Machida}, M.~N. 2024, \apj, 963, 72,
  \dodoi{10.3847/1538-4357/ad19ce}

\bibitem[{{Pascucci} {et~al.}(2023){Pascucci}, {Cabrit}, {Edwards}, {Gorti},
  {Gressel}, \& {Suzuki}}]{pascucci23}
{Pascucci}, I., {Cabrit}, S., {Edwards}, S., {et~al.} 2023, in Astronomical
  Society of the Pacific Conference Series, Vol. 534, Protostars and Planets
  VII, ed. S.~{Inutsuka}, Y.~{Aikawa}, T.~{Muto}, K.~{Tomida}, \& M.~{Tamura},
  567

\bibitem[{{Pinte} {et~al.}(2016){Pinte}, {Dent}, {M{\'e}nard}, {Hales}, {Hill},
  {Cortes}, \& {de Gregorio-Monsalvo}}]{pinte16}
{Pinte}, C., {Dent}, W.~R.~F., {M{\'e}nard}, F., {et~al.} 2016, \apj, 816, 25,
  \dodoi{10.3847/0004-637X/816/1/25}

\bibitem[{{Pizzati} {et~al.}(2023){Pizzati}, {Rosotti}, \&
  {Tabone}}]{pizzati23}
{Pizzati}, E., {Rosotti}, G.~P., \& {Tabone}, B. 2023, \mnras, 524, 3184,
  \dodoi{10.1093/mnras/stad2057}

\bibitem[{{Sato} {et~al.}(2023){Sato}, {Tokuda}, {Machida}, {Tachihara},
  {Harada}, {Yamasaki}, {Hirano}, {Onishi}, \& {Matsushita}}]{sato23}
{Sato}, A., {Tokuda}, K., {Machida}, M.~N., {et~al.} 2023, \apj, 958, 102,
  \dodoi{10.3847/1538-4357/ad0132}

\bibitem[{{Sheehan} \& {Eisner}(2017)}]{sheehan17}
{Sheehan}, P.~D., \& {Eisner}, J.~A. 2017, \apjl, 840, L12,
  \dodoi{10.3847/2041-8213/aa6df8}

\bibitem[{{Sheehan} \& {Eisner}(2018)}]{sheehan18}
---. 2018, \apj, 857, 18, \dodoi{10.3847/1538-4357/aaae65}

\bibitem[{{Sheehan} {et~al.}(2020){Sheehan}, {Tobin}, {Federman}, {Megeath}, \&
  {Looney}}]{sheehan20}
{Sheehan}, P.~D., {Tobin}, J.~J., {Federman}, S., {Megeath}, S.~T., \&
  {Looney}, L.~W. 2020, \apj, 902, 141, \dodoi{10.3847/1538-4357/abbad5}

\bibitem[{{Shoshi} {et~al.}(2024){Shoshi}, {Harada}, {Tokuda}, {Kawasaki},
  {Yamasaki}, {Sato}, {Omura}, {Yamaguchi}, {Tachihara}, \&
  {Machida}}]{shoshi24}
{Shoshi}, A., {Harada}, N., {Tokuda}, K., {et~al.} 2024, \apj, 961, 228,
  \dodoi{10.3847/1538-4357/ad12b5}

\bibitem[{{Simon} {et~al.}(2013{\natexlab{a}}){Simon}, {Bai}, {Armitage},
  {Stone}, \& {Beckwith}}]{simon13b}
{Simon}, J.~B., {Bai}, X.-N., {Armitage}, P.~J., {Stone}, J.~M., \& {Beckwith},
  K. 2013{\natexlab{a}}, \apj, 775, 73, \dodoi{10.1088/0004-637X/775/1/73}

\bibitem[{{Simon} {et~al.}(2013{\natexlab{b}}){Simon}, {Bai}, {Stone},
  {Armitage}, \& {Beckwith}}]{simon13a}
{Simon}, J.~B., {Bai}, X.-N., {Stone}, J.~M., {Armitage}, P.~J., \& {Beckwith},
  K. 2013{\natexlab{b}}, \apj, 764, 66, \dodoi{10.1088/0004-637X/764/1/66}

\bibitem[{{Suzuki} {et~al.}(2016){Suzuki}, {Ogihara}, {Morbidelli}, {Crida}, \&
  {Guillot}}]{suzuki16}
{Suzuki}, T.~K., {Ogihara}, M., {Morbidelli}, A., {Crida}, A., \& {Guillot}, T.
  2016, \aap, 596, A74, \dodoi{10.1051/0004-6361/201628955}

\bibitem[{{Tabone} {et~al.}(2022{\natexlab{a}}){Tabone}, {Rosotti}, {Cridland},
  {Armitage}, \& {Lodato}}]{tabone22a}
{Tabone}, B., {Rosotti}, G.~P., {Cridland}, A.~J., {Armitage}, P.~J., \&
  {Lodato}, G. 2022{\natexlab{a}}, \mnras, 512, 2290,
  \dodoi{10.1093/mnras/stab3442}

\bibitem[{{Tabone} {et~al.}(2022{\natexlab{b}}){Tabone}, {Rosotti}, {Lodato},
  {Armitage}, {Cridland}, \& {van Dishoeck}}]{tabone22b}
{Tabone}, B., {Rosotti}, G.~P., {Lodato}, G., {et~al.} 2022{\natexlab{b}},
  \mnras, 512, L74, \dodoi{10.1093/mnrasl/slab124}

\bibitem[{{Tabone} {et~al.}(2017){Tabone}, {Cabrit}, {Bianchi}, {Ferreira},
  {Pineau des For{\^e}ts}, {Codella}, {Gusdorf}, {Gueth}, {Podio}, \&
  {Chapillon}}]{tabone17}
{Tabone}, B., {Cabrit}, S., {Bianchi}, E., {et~al.} 2017, \aap, 607, L6,
  \dodoi{10.1051/0004-6361/201731691}

\bibitem[{{Tomida} {et~al.}(2017){Tomida}, {Machida}, {Hosokawa}, {Sakurai}, \&
  {Lin}}]{tomida17}
{Tomida}, K., {Machida}, M.~N., {Hosokawa}, T., {Sakurai}, Y., \& {Lin}, C.~H.
  2017, \apjl, 835, L11, \dodoi{10.3847/2041-8213/835/1/L11}

\bibitem[{{Tomida} {et~al.}(2013){Tomida}, {Tomisaka}, {Matsumoto}, {Hori},
  {Okuzumi}, {Machida}, \& {Saigo}}]{tomida13}
{Tomida}, K., {Tomisaka}, K., {Matsumoto}, T., {et~al.} 2013, \apj, 763, 6,
  \dodoi{10.1088/0004-637X/763/1/6}

\bibitem[{{Tomisaka}(1998)}]{tomisaka98}
{Tomisaka}, K. 1998, \apjl, 502, L163, \dodoi{10.1086/311504}

\bibitem[{{Tomisaka}(2002)}]{tomisaka02}
---. 2002, \apj, 575, 306, \dodoi{10.1086/341133}

\bibitem[{{Tsukamoto} {et~al.}(2022){Tsukamoto}, {Maury}, {Commer{\c{c}}on},
  {Alves}, {Cox}, {Sakai}, {Ray}, {Zhao}, \& {Machida}}]{tsukamoto22}
{Tsukamoto}, Y., {Maury}, A., {Commer{\c{c}}on}, B., {et~al.} 2022, arXiv
  e-prints, arXiv:2209.13765, \dodoi{10.48550/arXiv.2209.13765}

\bibitem[{{Turner} \& {Sano}(2008)}]{turner08}
{Turner}, N.~J., \& {Sano}, T. 2008, \apjl, 679, L131, \dodoi{10.1086/589540}

\bibitem[{{Uchida} \& {Shibata}(1985)}]{uchida85}
{Uchida}, Y., \& {Shibata}, K. 1985, \pasj, 37, 515

\bibitem[{{Vaytet} {et~al.}(2018){Vaytet}, {Commer{\c{c}}on}, {Masson},
  {Gonz{\'a}lez}, \& {Chabrier}}]{vaytet18}
{Vaytet}, N., {Commer{\c{c}}on}, B., {Masson}, J., {Gonz{\'a}lez}, M., \&
  {Chabrier}, G. 2018, \aap, 615, A5, \dodoi{10.1051/0004-6361/201732075}

\bibitem[{{Vorobyov} \& {Basu}(2005)}]{vorobyov05}
{Vorobyov}, E.~I., \& {Basu}, S. 2005, \apjl, 633, L137, \dodoi{10.1086/498303}

\bibitem[{{Vorobyov} \& {Basu}(2006)}]{vorobyov06}
---. 2006, \apj, 650, 956, \dodoi{10.1086/507320}

\bibitem[{{Williams} \& {Cieza}(2011)}]{williams11}
{Williams}, J.~P., \& {Cieza}, L.~A. 2011, \araa, 49, 67,
  \dodoi{10.1146/annurev-astro-081710-102548}

\bibitem[{{Yen} {et~al.}(2014){Yen}, {Takakuwa}, {Ohashi}, {Aikawa}, {Aso},
  {Koyamatsu}, {Machida}, {Saigo}, {Saito}, {Tomida}, \& {Tomisaka}}]{yen14}
{Yen}, H.-W., {Takakuwa}, S., {Ohashi}, N., {et~al.} 2014, \apj, 793, 1,
  \dodoi{10.1088/0004-637X/793/1/1}

\bibitem[{{Zapata} {et~al.}(2015){Zapata}, {Lizano}, {Rodr{\'\i}guez}, {Ho},
  {Loinard}, {Fern{\'a}ndez-L{\'o}pez}, \& {Tafoya}}]{zapata15}
{Zapata}, L.~A., {Lizano}, S., {Rodr{\'\i}guez}, L.~F., {et~al.} 2015, \apj,
  798, 131, \dodoi{10.1088/0004-637X/798/2/131}

\end{thebibliography}
\bibliographystyle{aasjournal}

\end{document}